\documentclass[useAMS,usenatbib]{mn2e}
\usepackage{amsmath}
\usepackage{amsfonts}
\usepackage{amssymb,graphicx}
\usepackage{lscape}
\usepackage[lofdepth,lotdepth]{subfig}
\usepackage{graphicx}
\usepackage{pdflscape}
\usepackage{booktabs}

\def\jnlref#1{{\rm#1}}
\def\aj{\jnlref{AJ}}
\def\araa{\jnlref{ARA\&A}}
\def\apj{\jnlref{ApJ}}
\def\apjl{\jnlref{ApJ}}
\def\apjs{\jnlref{ApJS}}

\def\apss{\jnlref{Ap\&SS}}
\def\aap{\jnlref{A\&A}}

\def\mnras{\jnlref{MNRAS}}
\def\nat{\jnlref{Nature}}

\def\pasj{\jnlref{PASJ}}

\def\nar{\jnlref{NewAR}}

\title[Deep \emph{Chandra} observations of the NGC\,4472 globular cluster black hole XMMU\,122939.7+075333]{Deep \emph{Chandra} observations of the NGC\,4472 globular cluster black hole XMMU\,122939.7+075333: Short term variability from the first globular cluster black hole binary}
\author[T. D. Joseph, T. J. Maccarone, R. P. Kraft and G. R. Sivakoff] {T. D. Joseph$^{1,2,4}$\thanks{E-mail:
tana@ast.uct.ac.za; thomas.maccarone@ttu.edu; kraft@head.cfa.harvard.edu; sivakoff@ualberta.ca}, T. J. Maccarone$^{3,2}$\footnotemark[1], R. P. Kraft$^{4}$\footnotemark[1] and G. R. Sivakoff$^{5}$\footnotemark[1]\\
$^{1}$University of Cape Town, Rondebosch, 7701, Republic of South Africa\\
$^{2}$University of Southampton, Southampton, SO17 1BJ, United Kingdom\\
$^{3}$Texas Tech University, Lubbock, TX, USA \\
$^{4}$Harvard-Smithsonian Center for Astrophysics, 60 Garden Street, MS-4, Cambridge, MA 02138, USA\\
$^{5}$Department of Physics, University of Alberta, CCIS 4-183, Edmonton, Alberta T6G 2E1, Canada}

\begin{document}

\date{Accepted ???. Received ???; in original form ???}

\pagerange{\pageref{firstpage}--\pageref{lastpage}} \pubyear{???}

\maketitle

\label{firstpage}

\begin{abstract}
In this paper we discuss the luminosity modulations and spectral analysis results of the recent deep observations of  XMMU\,122939.7+075333, the first black hole discovered in a globular cluster. The source has been detected many times, typically with L$_X>10^{39}$\,erg\,s$^{-1}$, but in a 2010 observation had faded to $L_X\sim 10^{38}$\,erg\,s$^{-1}$. In our 2011 observations, it has rebrightened to L$_X\sim 2\times10^{39}$\,erg\,s$^{-1}$. This significant increase in luminosity over a a relatively short time period is not consistent with the idea that the long term flux modulations displayed by XMMU\,122939.7+075333 are caused by the Kozai mechanism alone as had previously been suggested. Instead, given that the source shows "spiky" behaviour in its light curve, it seems likely that the faintness in 2010 was a result of a short observation that did not catch any bright epochs. We also find that when the source is brighter than average, it has an excess of soft ($<$0.7\,keV) photons. The spectral analysis reveals strong, albeit model-dependent, evidence of emission from highly ionised oxygen (O\,{\sc viii}) when the source is brighter than average. 
\end{abstract}


\section{Introduction}

XMMU122939.7+075333, which is coincident with a spectrally confirmed globular cluster (RZ\,2109) in the massive Virgo elliptical galaxy NGC4472 \citep{2001AJ....121..210R}, was the first X-ray emitting globular cluster identified to contain a black hole (BH) system, based on data from X-ray Multiple Mirror-Newton (XMM) satellite \citep{2007Natur.445..183M}. The peak X-ray luminosity of the source at discovery, L$_X=4\times10^{39}$\,erg\,s$^{-1}$, is more than 10 times the Eddington luminosity (L$_{\rm Edd}$) of a neutron star. This high luminosity makes the source an ultraluminous X-ray source (ULX; a non-nuclear X-ray source with L$_{\rm X} > 1\times 10^{39}$\,erg\,s$^{-1}$, assuming isotropic emission). {bf Moreover, its luminosity varied by a factor of seven in just a few hours, thereby ruling out the possibility that the system is a superposition of neutron stars and confirming it as a likely BH system.}

\par
\citet{2008MNRAS.386.2075S} carried out spectroscopic analysis of XMMU\,122939.7+075333 using \emph{XMM-Newton} data \citep{2007Natur.445..183M} and \emph{Chandra} data. They found that the spectra were well fit by a multicolour disk blackbody model and had an inner disk temperature of about \emph{k}T$_{\rm in}$= 0.15\,keV with the same underlying continuum whether the source was in a bright or faint phase, but with varying absorption column density. The results of this spectral analysis support the idea that the X-ray variability was due to a change in absorption column density. Further investigation by \citet{2010MNRAS.409L..84M} using four short \emph{Chandra} and \emph{Swift} observations of XMMU\,122939.7+075333 found that its luminosity appeared to be declining over a period of several years. 

\par 
XMMU\,122939.7+075333 has also been studied extensively at optical wavelengths. \citet{2007ApJ...669L..69Z} found [O\,{\sc iii}]\,$\lambda$\,5007 emission associated with the system. Optical spectroscopic follow up with higher signal-to-noise by \citet{2008ApJ...683L.139Z} revealed both [O\,{\sc iii}]\,$\lambda$\,5007 and [O\,{\sc iii}]\,$\lambda$\,4949 emission from XMMU\,122939.7+075333. In addition to the narrow (few hundred km\,s$^{-1}$) core already found by \citet{2007ApJ...669L..69Z}, these emission lines were found to have very broad wings with a velocity width of about 2000\,km\,s$^{-1}$ and were therefore blended. The total  emission line luminosity in the \citet{2008ApJ...683L.139Z} spectrum was found to be 1.4$\times10^{37}$\,erg\,s$^{-1}$, with a flux ratio of [O\,{\sc iii}] to H$\beta$ of at least 30. A long term spectroscopic campaign was carried out by \citet{2011ApJ...739...95S} to monitor the variability and velocity structure of the [O\,{\sc iii}] emission from XMMU\,122939.7+075333. While the [O\,{\sc iii}] velocity structure was relatively constant over a time span of more than 400 days, the luminosity of the emission lines declined by approximately 10\%. Most recently, \citet{2012ApJ...759..126P}  used spatially resolved spectroscopic data from \emph{HST}/STIS to determine the spatial structure of the [O\,{\sc iii}] emission. They found that the half-light radius of the gas is between 3 and 7\,pc and that the light profile of the oxygen gas emission is consistent with the light profile of RZ\,2109's continuum.

\par
The high ratio of [O\,{\sc iii}] to H$\beta$ determined by \citet{2008ApJ...683L.139Z} suggests that the donor star is likely to be a hydrogen deficient star such as a white dwarf (WD), as first suggested by \citet{2009ApJ...705L.168G}. Most WDs are composed almost entirely ($>99\%$ by mass) of carbon and oxygen \citep[e.g.,][]{2011wdac.book...53D}. Thus we would expect that the matter accreted from such a donor would be rich in carbon and oxygen and poor in hydrogen. This is in keeping with the findings of \citet{2008ApJ...683L.139Z}, who showed that the optical spectrum associated with XMMU\,122939.7+075333 is poor in hydrogen and rich in oxygen. 

\par
On the basis of evolutionary arguments \citet{2010ApJ...717..948I} suggested that the system is most likely a hierarchical triple star system, with an inner stellar mass BH-WD binary. In this scenario, it is thought that the outer star in the triple plays an important role in the evolution of the inner binary through the Kozai mechanism \citep{1962AJ.....67..591K}.  \citet{2010MNRAS.409L..84M} argued that we would expect $\dot{m}$ (and therefore the X-ray flux of the binary) to vary significantly on the time-scale for which $e$ varies due to the Kozai mechanism. They showed that for representative values of the system parameters, the period of $e$ variations is about 40 years; we then expect the X-ray flux of the inner binary to vary on a similar timescale. \citet{2010MNRAS.409L..84M} studied X-ray data for XMMU\,122939.7+075333 that spanned 30 years and they show that long term variation in flux is consistent with it declining on the timescale of the period of eccentricity. Thus the scenario of XMMU\,122939.7+075333 as a triple system with an inner WD-BH binary was favoured. 

\par
The X-ray variability is likely to be due to a variable absorption column density caused by an accretion disk which has the same composition as the WD donor star. The [O\,{\sc iii}] emission is likely to be from the disk wind, driven by moderately super-Eddington accretion onto the central compact object. The disk wind is then photoionised by the X-ray emission from the central source.

\par
Such a system would have several observational features and parameters that could be observed to test the veracity of this model. A WD-BH binary is expected to be compact and have a short orbital period \citep{2004ApJ...607L.119B}. This gives rise to a high accretion rate, a fully ionised accretion disk and thus a persistently luminous X-ray source. The WD would have supplied the accretion disk with material rich in oxygen and carbon. It may therefore be possible to detect X-ray and/ or ultraviolet emission lines from carbon or oxygen in the source spectra. If the photoionised accretion disk wind is variable, we expect to see aperiodic modulations of the X-ray luminosity due to obscuration by a changing absorption column density \citep{1998MNRAS.295..595P}. Since soft X-rays ($<$1\,keV) are absorbed preferentially, this means that the source spectrum should be softer when the source is more luminous.

\par
Recent deep observations of XMMU\,122939+075333 have made it possible to probe some of these questions. In this paper we explore both the temporal and spectral properties of this source to improve our understanding of its nature. 

\section{Observations and Data Analysis} 

The two most recent deep observations of XMMU\,122939+075333 were obtained from the \emph{Chandra} telescope (Obs ID 12889 and 12888). They were taken one week apart on 14 and 21 February 2011 respectively, with exposure times of 140\,ks and 160\,ks respectively.

\par
The data were first reprocessed using \emph{Chandra} calibration data from March 2011 to create new level two events files with the CIAO \emph{acis\_reprocess\_events} script. The images were then checked for background flares. We used \emph{specextract} to produce source and background spectra and \emph{dmextract} to make light curves. 

\par
To establish whether the source is persistently bright, we look to a long term X-ray light curve spanning several years. The long term light curve of the source was produced using data from the \emph{Chandra}, \emph{XMM} and \emph{ROSAT} telescopes. The light curves for Obs ID 12889 and 12888 are used to explore the short term variation of the X-ray flux. 

\par
To test whether varying intrinsic absorption plays a significant role in flux variation, the most recent \emph{Chandra}observations were split into bright and faint phases (with the dividing line being the mean count rate of $\sim$1.3$\times10^{-3}$ to more accurately analyse the data. We analyse the light curves to determine whether the flux variations are between faint and bright phases is significant and, if so, whether it is related to a change in absorption column by calculating hardness ratios. 

\par
To do this, the counts in the bright and faint phases were divided into hard (1$-$5\,keV)  and soft (0.5$-$1\,keV) energy bands, such that there were a roughly even number of counts in each band for the bright phase. XMMU122939+075333 is situated near the edge of CCD chip ACIS-5 in Obs ID 12889 and 12888. To calculate the source count rate accurately, we made exposure maps of the chip using \emph{mkexpmap} and used them to correct for the decreased effective area near the chip gaps. The hardness ratio, $HR= (H-S)/(H+S)$ where $H$ and $S$ are the number of counts in the hard and soft bands respectively, was then determined for the bright and soft phase of each observation.  We also compute the probability that the changes in hardness ratio between faint and bright phases happened by chance, using the binomial probability distribution.

\par
In order to compare the spectral properties of XMMU\,122939+075333 to those of other black hole binaries, we model the spectra with the standard spectral model often used for Galactic BH binaries as well as ULXs; i.e. a model consisting of a disk blackbody (diskbb) continuum plus a powerlaw (PL) component. The Obs ID 12888 and 12889 spectra were both fit with this model. We also include 100\,ks \emph{XMM} observation of the source taken in 2004 \citep[hereafter referred to as XMM04;][]{2007Natur.445..183M} in out spectral analysis. 

\par
The optical spectra of XMMU\,122939+075333 shows strong [O\,{\sc iii}] emission. To establish whether oxygen emission is also present in the X-ray spectra of the source, we attempt to model the excess of soft counts as a an emission line, using a Gaussian spectral model component. The Obs ID 12889 and 12888 spectra and XMM04 were fit jointly with the same models so that the Gaussian component could be tied to the same value across all three data sets. The data sets were fit simultaneously with the same continuum model (either diskbb only, PL only, or diskbb plus PL), with the associated continuum spectral parameters (e.g., photon index, $\Gamma$) free to vary between each data set. The \emph{F} test was carried out on the Gaussian line component to determine its significance. 

\par
The spectral analysis was carried out using the Interactive Spectral Interpretation System (ISIS) version 1.6.1. Only data in the well calibrated, low background range of 0.5$-$5\,keV were used. This energy range is not standard; many other groups use an energy range up to 8 or 10\,keV. However, after inspecting our data, we found that our chosen energy range resulted in the best quality spectra. The Gehrels statistic was used to calculate errors as it gives a more reliable fit to data with low count rates \citep{1986ApJ...303..336G}. The phabs absorption model was used to fit foreground absorption with the Galactic absorption column density, $N_{\rm H}=1.6\times10^{20}$\,cm$^{-2}$ \citep{1990ARA&A..28..215D} and default ISIS (Xspec) abundances and cross sections. We do not have a sufficient number of counts in the spectra to allow for the fitting of models where both the spectral parameters and the absorption column density are free to vary. We thus keep the absorption column density frozen to the Galactic foreground value. The interstellar medium of elliptical galaxies contains very little cool molecular or atomic gas \citep[see e.g.,][]{2010ApJ...725..100W}, so any absorption from NGC\,4472 itself would be negligible. 

\par
For the cases where the flux variation between bright and faint spectra was found to be significant, the spectra were analysed separately. The bright and faint phase spectra were grouped to a minimum signal-to-noise ratio of three and two per bin respectively. These spectra were then separately fit with PL or diskbb continuum models \citep{1989PASJ...41...97M} and a Gaussian component to fit an excess of counts at $\sim$0.6\,keV.

\par
To further test whether the flux variation between bright and faint spectra could be due to a change in absorption column density, we also fit the faint spectra with the best fit parameters of their respective bright spectrum and allow $N_{\rm H}$ to vary.

\section {Results} 

\subsection{Long term light curve}

The long term light curve was made up of data from the most recent deep \emph{Chandra} observations of XMMU\,122939.7+075333 as well as archival data from \emph{XMM}, \emph{Chandra}, and \emph{ROSAT}.

\par
The flux from a \emph{Chandra} data set taken in 2000 was determined using ISIS and the spectral parameters from \citet{2008MNRAS.386.2075S}. The flux for the 2004 \emph{XMM} observation was taken from the bright phase spectrum (count rate $\geqslant$ 0.04 counts/s) of XMM04 only and calculated in ISIS using the spectral parameters found in \citet{2008MNRAS.386.2075S}. This was done to be consistent with the work of \citet{2008MNRAS.386.2075S} and \citet{2010MNRAS.409L..84M}. The 2002 \emph{XMM} \citep[values taken from the 2\emph{XMM} catalogue of][]{2009A&A...493..339W} and 2010 \emph{Chandra} \citep{2010MNRAS.409L..84M} fluxes and errors were calculated from their respective count rate or flux errors using the {\sc w3pimms} tool, assuming a $\Gamma$= 1.7 power law spectral model. 

\par
The \emph{ROSAT} data \citep{2002ApJS..143...25C} had poor spectral resolution and {\sc w3pimms} was used to calculate the flux from the count rate. \citet{2008MNRAS.386.2075S} found that the count rate for the \emph{ROSAT} data was consistent with that of the bright XMM04 data; they therefore used the XMM04 inner disk temperature value (0.2\,keV) to fit the \emph{ROSAT} data.  The \emph{ROSAT} luminosity is a factor of 6 higher when a 1\,keV blackbody temperature is assumed compared to a temperature of  0.2\,keV. The errors for this source are thus dominated by the uncertain energy to counts conversion for this data. \citet{2010MNRAS.409L..84M} estimate that the luminosity of the source at the time of the \emph{ROSAT} observation is between 10$^{39}$ and 10$^{40}$ erg\,s$^{-1}$. In this work we used a 0.2\,keV blackbody to estimate the flux of the \emph{ROSAT} data.

\par
\begin{figure*}
\centering
\includegraphics[width=0.5\textwidth]{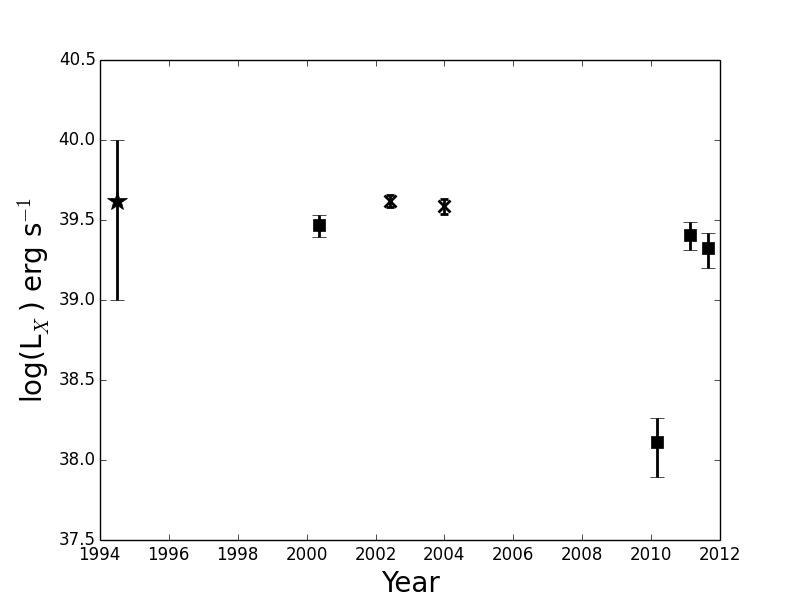}
\caption[The long term light curve of XMMU\,122939.7+075333]{The long term light curve of XMMU\,122939.7+075333. The squares, crosses, and star represent \emph{Chandra}, \emph{XMM}, and \emph{ROSAT} data respectively. All luminosities are in the 0.2--10\,keV range. The errors for the \emph{Chandra} fluxes (except those taken in 2010) as well as that of XMM04 reflect the 90\,\% confidence intervals. For clarity, the data point for Obs ID 12888 has been plotted with offsets of six months.  }
\label{long_lc}
\end{figure*}

All the data plotted in the light curve, shown in Fig.\,\ref{long_lc}, are the unabsorbed X-ray luminosities for the 0.2--10\,keV range. The \emph{ROSAT}, 2002 \emph{XMM} and 2010 \emph{Chandra} errors were calculated from their respective count rate or flux errors using the {\sc w3pimms} tool. The dominant source of error for the \emph{ROSAT} luminosity is due to uncertainty in converting counts to energy \citep[see][]{2008MNRAS.386.2075S, 2010MNRAS.409L..84M}. The errors in these luminosity measurements do not account for the uncertainty in the power law slope.

\subsection{Light curves of Obs ID 12889 and 12888}\label{short_lc_results}

The source displays periods of brightening (by a factor of roughly three from the mean count rate) on the timescale of hours in the light curves from Obs ID 12289 and 12888. These source light curves were fit with a constant count rate, which was different for each light curve. This yielded a $\chi^2/\nu$ of 35.95/34 and 68.86/40 for Obs ID 12889 and 12888 respectively.  A constant count rate fit to the light curve for  Obs ID 12889 cannot be ruled out at greater than the 37\% level; we did not detect significant variability over this observation. We will therefore not discuss the intra-observational variability of Obs ID 12889 any further in this paper. However, for Obs ID 12888, the straight line can be ruled out at greater than the 99\,\% level; the source is displaying significant flux variation over this time period.

\subsection{Hardness Ratios}

The count rate threshold between the bright and faint phases is 1.3$\times10^{-3}$\,counts/s. 

We use the binomial probability distribution to determine whether or not the change in HR between bright and faint phases is significant, with each photon simply being assigned a probability of being in the soft or hard band. In our case, the null hypothesis is that the HR of the faint phases is only different from that of the bright ones by chance (given that in the bright phases the chances of a photon falling into the hard or soft band is 50\%). Hardness ratio data as well as the probabilities are displayed in Table\,\ref{HR_table}.

\par
We find that the hardness ratios for the bright and faint phases of Obs ID 12888 are 0.02$\pm$0.11 and 0.38$\pm$0.13. The null hypothesis probability for Obs ID 12888 is 0.006; thus the difference in hardness between the faint and bright phases is statistically significant.

\begin{table*}
\centering
\begin{tabular}{|l|l|}
\toprule
Phase & HR  \\
\midrule
12888 bright &  0.02 $\pm$ 0.11   \\
12888 faint &  0.38 $\pm$ 0.13  \\
\midrule
Null hypothesis probability:& 0.006 \\
\bottomrule
\end{tabular}
\caption[Hardness ratio data for XMMU\,122939.7+075333]{ \footnotesize Hardness ratio data for XMMU\,122939.7+075333 for Obs ID 12888. The probability column indicates the null hypothesis probability that the faint phase HR could be produced from a spectrum that is the same as in the bright phase.}
\label{HR_table}
\end{table*}

\subsection{Spectral Analysis}

\subsubsection{Disk blackbody plus power law spectral model}\label{disk_PL}

\citet{2006MNRAS.371.1877R,2006MNRAS.368..397S} and \citet{2007Ap&SS.311..203R} showed that some of the ULX spectra they studied could be modelled using the standard disk blackbody plus power law model used to model the spectra of Galactic binaries, i.e., the soft end of the spectrum was fit by the diskbb component and the hard end was fit by the PL. However, \citet{2006MNRAS.368..397S} also found that a significant proportion of their sources had spectra that were better fit with an ``inverted'' model, i.e., a steep PL component fit to the soft end of the spectrum and diskbb component fit for the harder end. In this section we explore whether the disk plus power law model provides a good fit to the spectra of XMMU\,122939.7+075333.

\paragraph{Bright phase spectrum } 
We are unable to fit complex physical models to the bright and faint phase spectra of Obs ID 12888 due to the low number of counts (66 total counts for the faint phase spectrum of Obs ID 12888). The following spectral fits are used only to parametrise the data. In these spectral models, the absorption column density is frozen to the Galactic foreground value \citep[$N_{\rm H}=1.6\times10^{20}$\,cm$^{-2}$;][]{1990ARA&A..28..215D}. 

\par
The bright phase of Obs ID 12888 was individually fit with a diskbb plus PL model. For Obs ID 12888, the disk component has an inner temperature, $0.08^{+0.06}_{-0.01}$\,keV and the PL component is $\Gamma=2.06^{+1.50}_{-1.42}$, producing an acceptable fit with $\chi^2$/$\nu=4.49/3$.


\begin{figure*}
\centering
\subfloat{\includegraphics[width=0.43\textwidth]{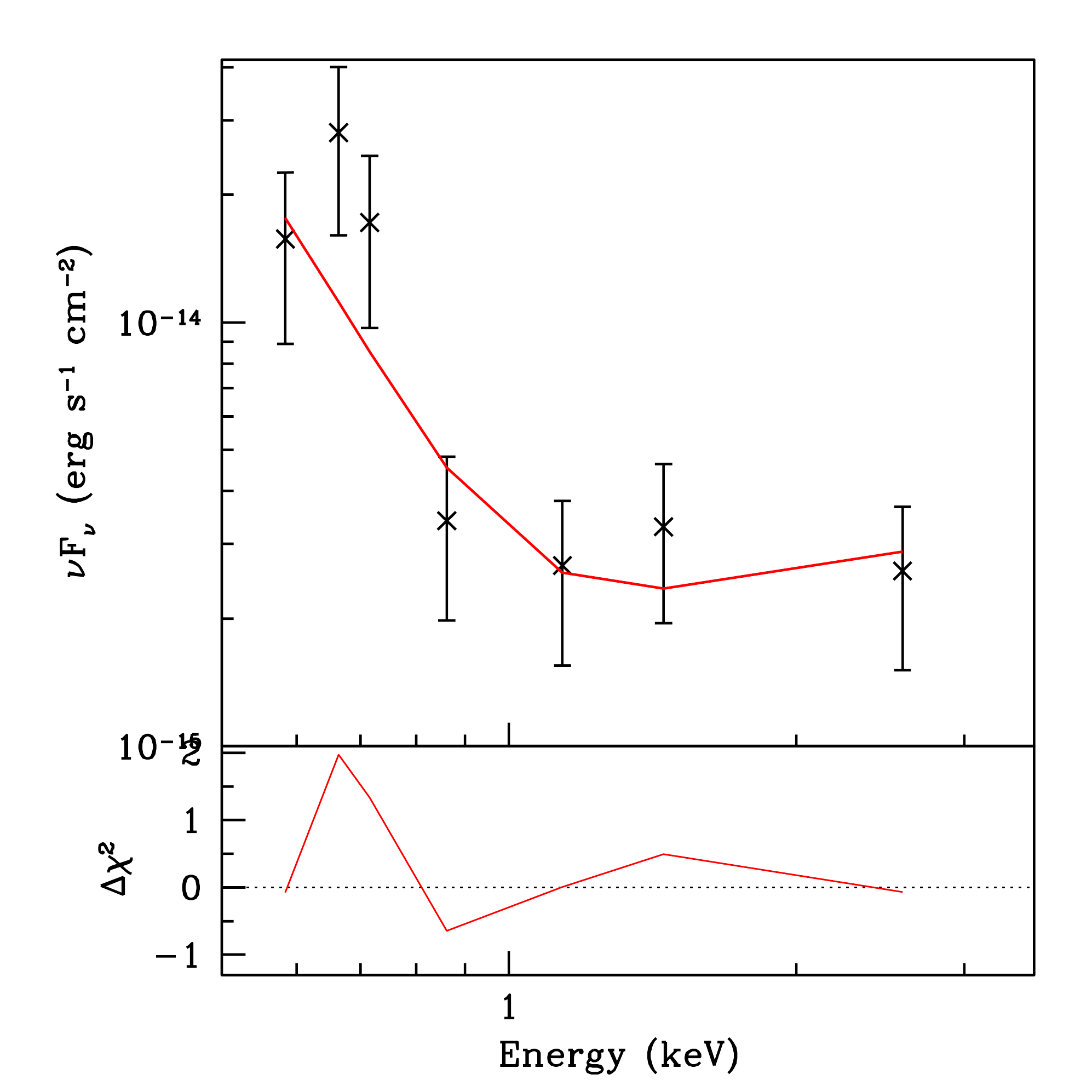}}
\caption[The bright unfolded spectrum of XMMU\,122939.7+075333 fit with diskbb plus PL model taken from Obs ID 1288]{The bright unfolded spectrum of XMMU\,122939.7+075333 taken from Obs ID 12888 with its respective best fit diskbb plus PL model (red line; see Table\,\ref{disk_PL_params} for spectral model fit parameters). The $\Delta$$\chi^2$ residual values are shown below the spectrum. }
\label{bright_diskPL_spectrum}
\end{figure*}

\begin{table*} 
\centering
\scriptsize
\begin{tabular}{lllllll}
\toprule
Spectrum & Counts & $\chi^2$/$\nu$ &  Diskbb norm. &T$_{\rm in}$ & PL norm. & $\Gamma$ \\
  &  &  & & (keV) & ($\times10^{-6}$) & \\
\midrule
Obs ID 12888 bright & 107 & 4.49/3 & $346^{+1.3\times10^8}_{-342}$ & $0.08^{+0.06}_{-0.01}$ & $1.6^{+1.1}_{-1.2}$ &  $2.06^{+1.50}_{-1.42}$ \\
Obs ID 12889 & 242 & 9.99/28 & 29.0$^{+24085}_{-28.3}$ & $0.12^{+0.03}_{-0.06}$ & 2.76$^{+4.50}_{-2.28}$ & $1.28^{+1.20}_{-0.96}$  \\
Obs ID 12888 & 173 & 8.51/19  & 225.2$^{+5644}_{-222.1}$ & $0.09^{+0.02}_{-0.04}$ & 1.99$^{+1.69}_{-1.32}$ & $1.45^{+1.29}_{-1.06}$  \\
XMM04 & 2532 & 61.1/81 & 0.55$^{+3.94}_{-0.30}$ & $0.26^{+0.06}_{-0.07}$ & 1.79$^{+1.9\times10^7}_{-1.72}$ & $0.60^{+1.92}_{->2.60}$   \\
\bottomrule
\end{tabular}
\caption[The best fit spectral parameters for the diskbb + PL spectral model]{The best fit spectral parameters for the  diskbb plus PL spectral model. The errors shown are the 90\% confidence intervals.}
\label{disk_PL_params}
\end{table*}

\paragraph{Obs ID 12889, 12888, and XMM04 full spectra}

\begin{figure*}
 \centering
\subfloat{\includegraphics[width=0.45\textwidth]{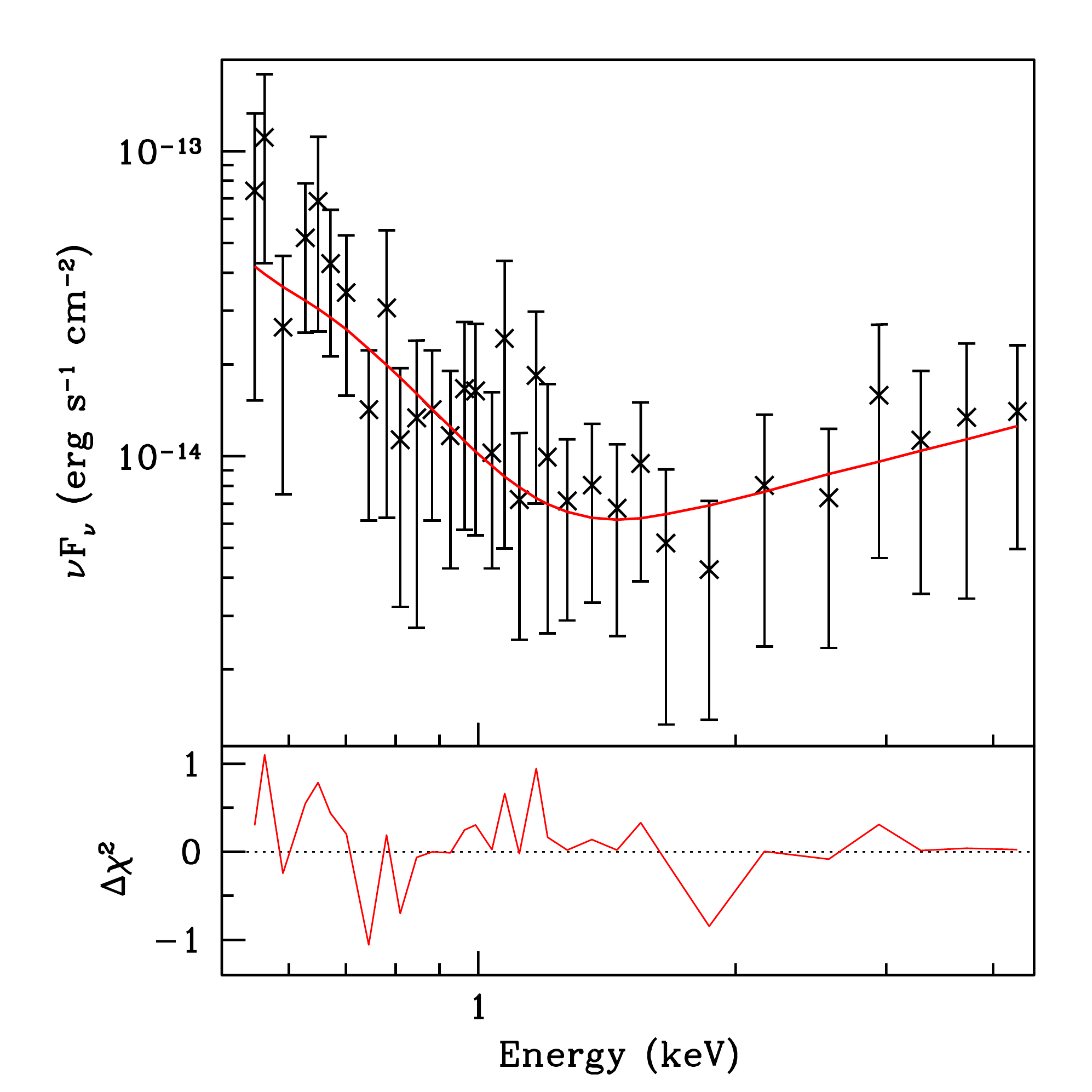} }
\qquad
\subfloat{\includegraphics[width=0.45\textwidth]{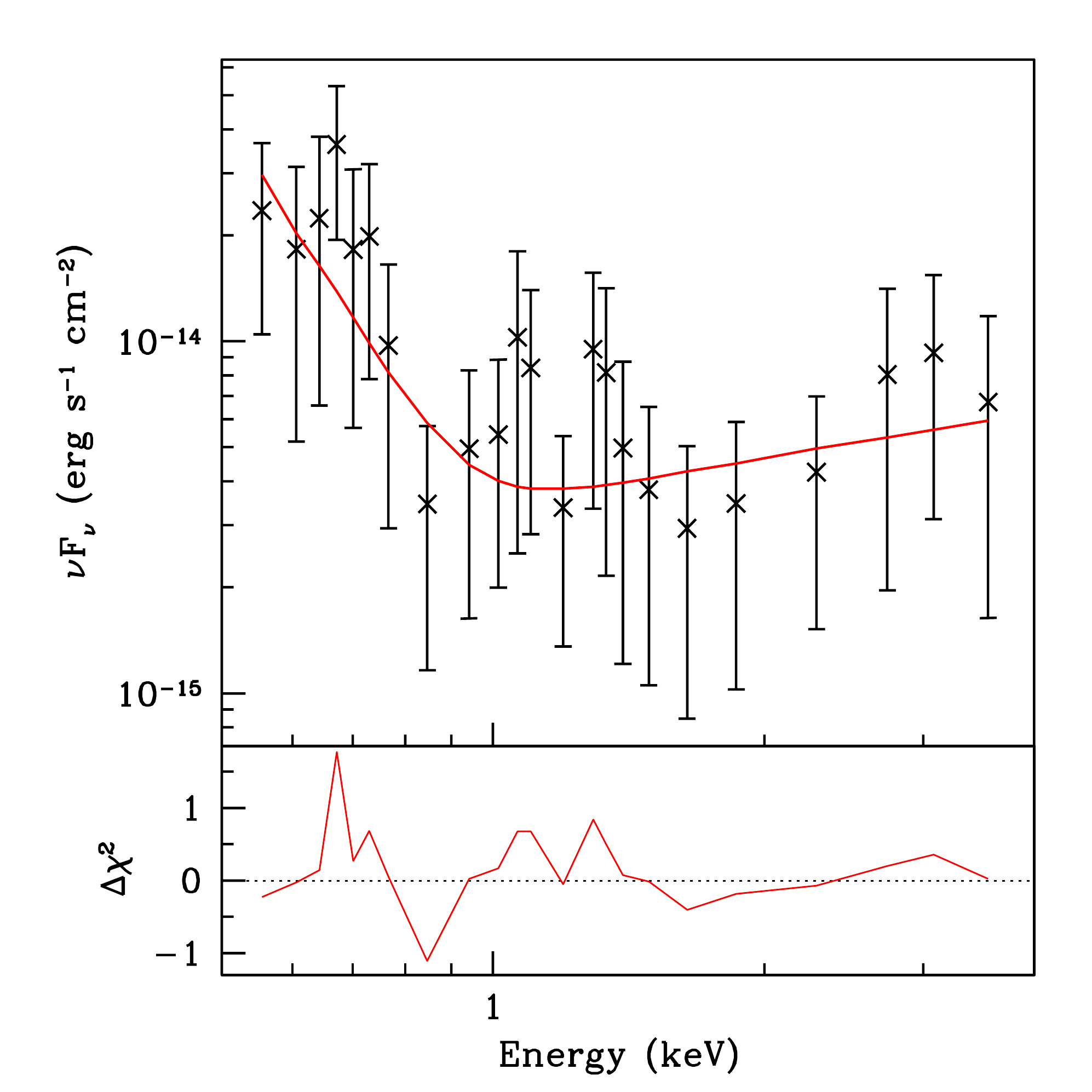} }
\qquad
\subfloat{\includegraphics[width=0.45\textwidth]{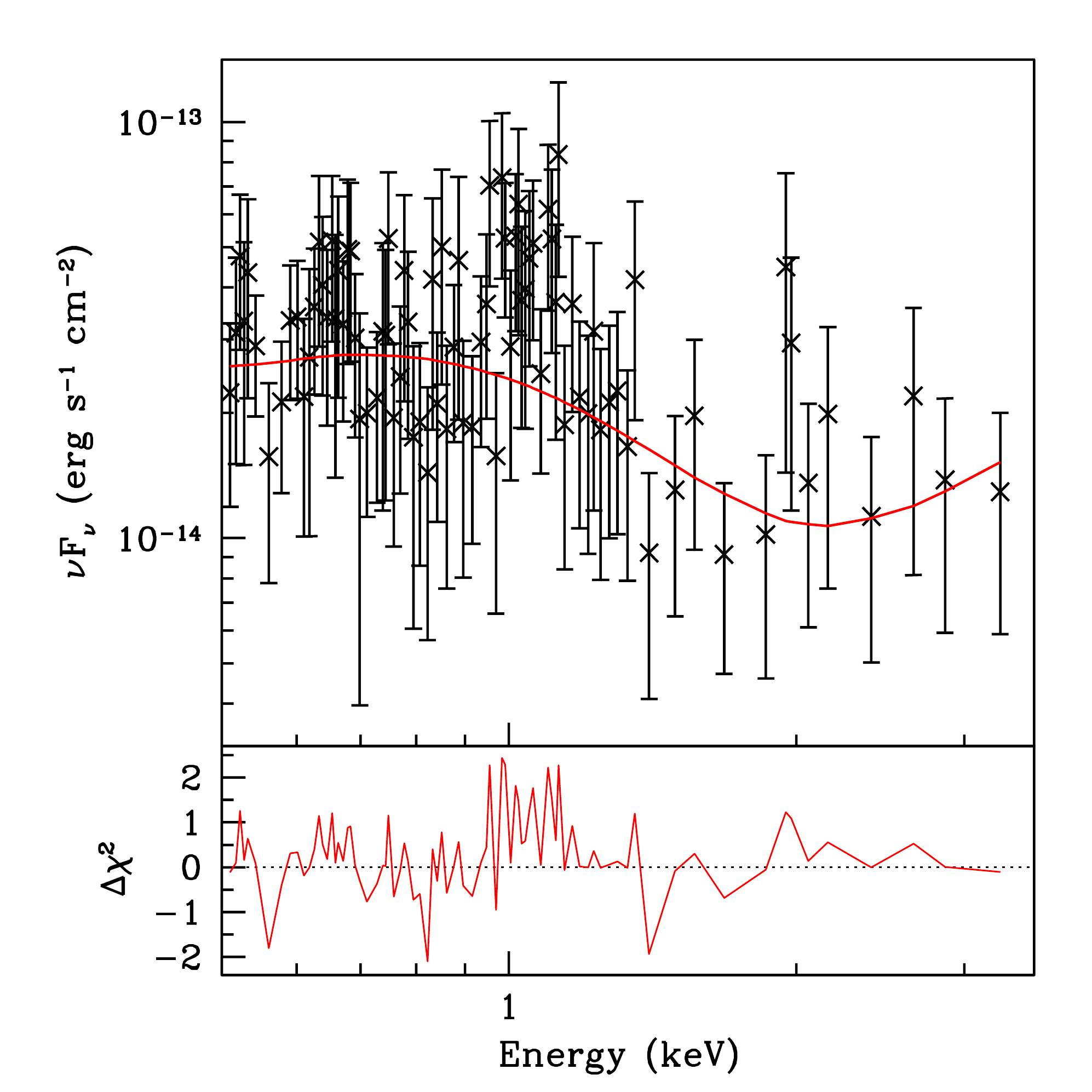} }
\caption[The unfolded spectra of XMMU\,122939.7+075333 fit with a diskbb plus PL model]{The unfolded spectra of XMMU\,122939.7+075333 taken from Obs ID 12889 (top left panel) and 12888 (top right panel) and XMM04 (bottom panel) with a diskbb plus PL model (red line; see Table\,\ref{disk_PL_params} for spectral model fit parameters). The $\Delta$$\chi^2$ residual values are shown below each spectrum. }
\label{disk_PL_spectra}
\end{figure*}

The disk blackbody plus power law model produced acceptable fits for all three data sets. We note that for our spectra the PL slopes are not well constrained, especially for XMM04. Nevertheless, all three sets of spectral fit parameters were consistent with the source having a cool disk and hard PL component. T$_{\rm in}$ was found to be $0.09^{+0.02}_{-0.04}$, $0.12^{+0.03}_{-0.06}$ and  $0.26^{+0.06}_{-0.07}$ for Obs ID 12888, 12889 and XMM04 respectively. The PL slopes were $\Gamma=1.45^{+1.29}_{-1.06}$, $\Gamma=1.28^{+1.20}_{-0.96}$ and $\Gamma=0.60^{+1.92}_{->2.60}$ for Obs ID 12888, 12889 and XMM04 respectively (the $>$ symbol indicates that the spectral parameter pins at -2). These parameters are summarised in Table\,\ref{disk_PL_params} and the spectra are shown in Fig.\,\ref{disk_PL_spectra}.

\subsubsection{Continuum plus Gaussian spectral model}

\paragraph{Continuum in bright and faint phases}

The bright and faint phase continua of Obs ID 12888 were better fit by the diskbb model with inner disk temperature, $k$T$_{in}$=0.6$^{+0.1}_{-0.2}$ and 0.5$^{+1.2}_{-0.2}$\,keV respectively.

\paragraph{Gaussian component in bright and faint phases}
The bright phase spectrum of Obs ID 12888 is better fit when a Gaussian component is included in the spectral model to account for the excess of soft photons at approximately 0.6\,keV. 
For Obs ID 12888 bright phase the line energy is 0.65$^{+0.04}_{-0.2}$ with a normalisation of 6.4$\times10^{-6}$ $^{+2.2}_{-3.4\times10^{-6}}$. 
The physical width (PW) and equivalent width (EW) for the bright Obs ID 12888 spectrum are 0.1$^{+0.1}_{-0.1}$\,keV and 2.1$^{+7.2\times10^5}_{-1.0}$\,keV respectively. The line luminosity is $\sim$10$^{38}$\,erg\,s$^{-1}$. The faint phase spectrum did not require the Gaussian component in order to have acceptable fits.

\begin{table*}
\centering
\tiny
\begin{tabular}{lllllllllll}
\toprule
Spectrum & Counts & $\chi^2$/$\nu$ & Diskbb norm. & T$_{\rm in}$ & PL norm. & $\Gamma$  & Gauss. norm. & Line energy & Physical width & Eq. width  \\
  &  &  & & (keV) & ($\times10^{-6}$)  & & ($\times10^{-6}$) & (keV) &  (keV)  & (keV)  \\
\midrule
12888 faint & 66 & 0.48/3 & 0.004$^{+2.6}_{-0.004}$ &  0.5$^{+1.2}_{-0.2}$ & -- &  -- & -- & -- & --& --   \\
12888 bright & 107  & 0.10/2 & 0.003$^{+1.4}_{-2.6\times10^{-3}}$ &  0.6$^{+0.1}_{-0.2}$ & -- & -- & 6.4$^{+2.1\times10^6}_{-3.4}$ & 0.65$^{+0.04}_{-0.2}$ & 0.1$^{+0.1}_{-0.1}$ & 2.1$^{+7.2\times10^5}_{-1.0}$  \\
Obs ID 12889 & 242 & 118.4/134 & 0.4$^{+2.2}_{-0.3}$ & 0.3$^{+0.1}_{-0.1}$ & -- & -- & -- & -- & -- & -- \\
Obs ID 12888 & 173 & 118.4/134 & 0.01$^{+1.5}_{-0.01}$ & 0.5$^{+1.4}_{-0.2}$ & -- & -- & --  & -- & -- & -- \\
XMM04 & 2532 & 118.4/134 & 0.3$^{+0.2}_{-0.1}$ & 0.3$^{+0.04}_{-0.04}$ & -- & -- & -- & -- & -- & -- \\ 
Obs ID 12889 & 242 & 99.6/131 & 0.2$^{+1.0}_{-0.2}$ & 0.3$^{+0.2}_{-0.1}$ & -- & -- & 3.4$^{+1.9\times10^7}_{-1.3}$ & 0.66$^{+0.01}_{-0.04}$ & 6.2$^{+4000}_{-6.2}$$\times10^{-5}$ & 0.9$^{+4.8\times10^6}_{-0.3}$ \\
Obs ID 12888 & 173 & 99.6/131 & 1.6$^{+1480}_{-1.62}$$\times10^{-3}$ & 0.7$^{+2.4}_{-0.4}$ & -- & -- & 3.4$^{+1.9\times10^7}_{-1.3}$ & 0.66$^{+0.01}_{-0.04}$ & 6.2$^{+4000}_{-6.2}$$\times10^{-5}$ & 0.2$^{+9.5\times10^5}_{-0.02}$\\
XMM04 & 2532 & 99.6/131 & 0.1$^{+0.1}_{-0.1}$ & 0.4$^{+0.1}_{-0.1}$ &  -- & -- & 3.4$^{+1.9\times10^7}_{-1.3}$ & 0.66$^{+0.01}_{-0.04}$ & 6.2$^{+4000}_{-6.2}$$\times10^{-5}$ & 0.1$^{+5.5\times10^5}_{-0.04}$ \\
Obs ID 12889 & 242 & 98.2/134 & -- & -- & 8.3$^{+1.8}_{-1.9}$ & 3.5$^{+0.9}_{-0.6}$ & -- &-- & -- & -- \\
Obs ID 12888 & 173 & 98.2/134 & -- & -- & 3.7$^{+1.2}_{-1.3}$ & 2.8$^{+1.5}_{-1.1}$ & -- &-- & -- & -- \\
XMM04 & 2532 & 98.2/134 & -- & -- & 14.1$^{+1.3}_{-1.3}$ & 2.7$^{+0.2}_{-0.2}$ & -- & -- & -- & -- \\
Obs ID 12889 & 242 & 86.5/131 & -- & -- & 7.7$^{+1.9}_{-2.2}$ & 3.1$^{+0.9}_{-0.8}$ & 2.7$^{1.4}_{-1.3}$ & 0.66$^{+0.02}_{-0.02}$ & 1.9$^{+40000}_{-1.92}$$\times10^{-6}$ & 0.3$^{+0.2}_{-0.2}$ \\
Obs ID 12888 & 173 & 86.5/131 & -- & -- & 3.2$^{+1.2}_{-1.2}$ & 2.3$^{+1.1}_{-0.8}$ & 2.7$^{1.4}_{-1.3}$ & 0.66$^{+0.02}_{-0.02}$ & 1.9$^{+40000}_{-1.92}$$\times10^{-6}$ & 0.1$^{+0.05}_{-0.05}$ \\
XMM04 & 2532 & 86.5/131 & -- & -- & 13.0$^{+1.4}_{-1.4}$ & 2.5$^{+0.2}_{-0.2}$ & 2.7$^{1.4}_{-1.3}$ & 0.66$^{+0.02}_{-0.02}$ & 1.9$^{+40000}_{-1.92}$$\times10^{-6}$ & 0.1$^{+0.01}_{-0.01}$ \\
Obd ID 12889 & 242 & 71.4/125 & 2.4$^{+3.7\times10^{13}}_{-1.4}$$\times10^{-12}$ & 997$^{+>3.0}_{-996}$ & 5.6$^{+>10^{16}}_{-5.6}$ & 4.8$^{+>2.2}_{->6.8}$& 2.3$^{+1.4}_{-1.5}$ & 0.66$^{+0.02}_{-0.02}$ & 0.01$^{+0.03}_{->0.0}$ & 0.1$^{+0.03}_{-0.03}$\\
Obs ID 12888 & 173 & 71.4/125 & 214$^{+0.1}_{-213.9}$ & 0.1$^{+0.03}_{-0.05}$ & 2.2$^{+0.8}_{-2.1}$ & 1.6$^{+0.2}_{-1.5}$ & 2.3$^{+1.4}_{-1.5}$ & 0.66$^{+0.02}_{-0.02}$ & 0.01$^{+0.03}_{->0.0}$ & 0.1$^{+0.07}_{-0.07}$\\
XMM04 & 2532 & 71.4/125 & 0.3$^{+0.2}_{-0.2}$ & 0.3$^{+0.003}_{-0.01}$ & 1.0$^{+3.7\times10^7}_{-0.01}$ & 0.2$^{+2.1}_{-1.1}$ & 2.3$^{+1.4}_{-1.5}$ & 0.66$^{+0.02}_{-0.02}$ & 0.01$^{+0.03}_{->0.0}$ & 0.1$^{+0.04}_{-0.04}$\\
\bottomrule
\end{tabular}
\caption[The best fit spectral parameters for the continuum only and continuum + Gaussian spectral models]{The best fit spectral parameters for the continuum only and continuum plus Gaussian spectral models. The errors shown are the 90\% confidence intervals. For the joint spectral fit with diskbb continuum, the physical width of the Gaussian component was restricted to 0$-$0.04\,keV.}
\label{joint_fit}
\end{table*}

\begin{figure*}
\centering
\subfloat{\includegraphics[width=0.45\textwidth]{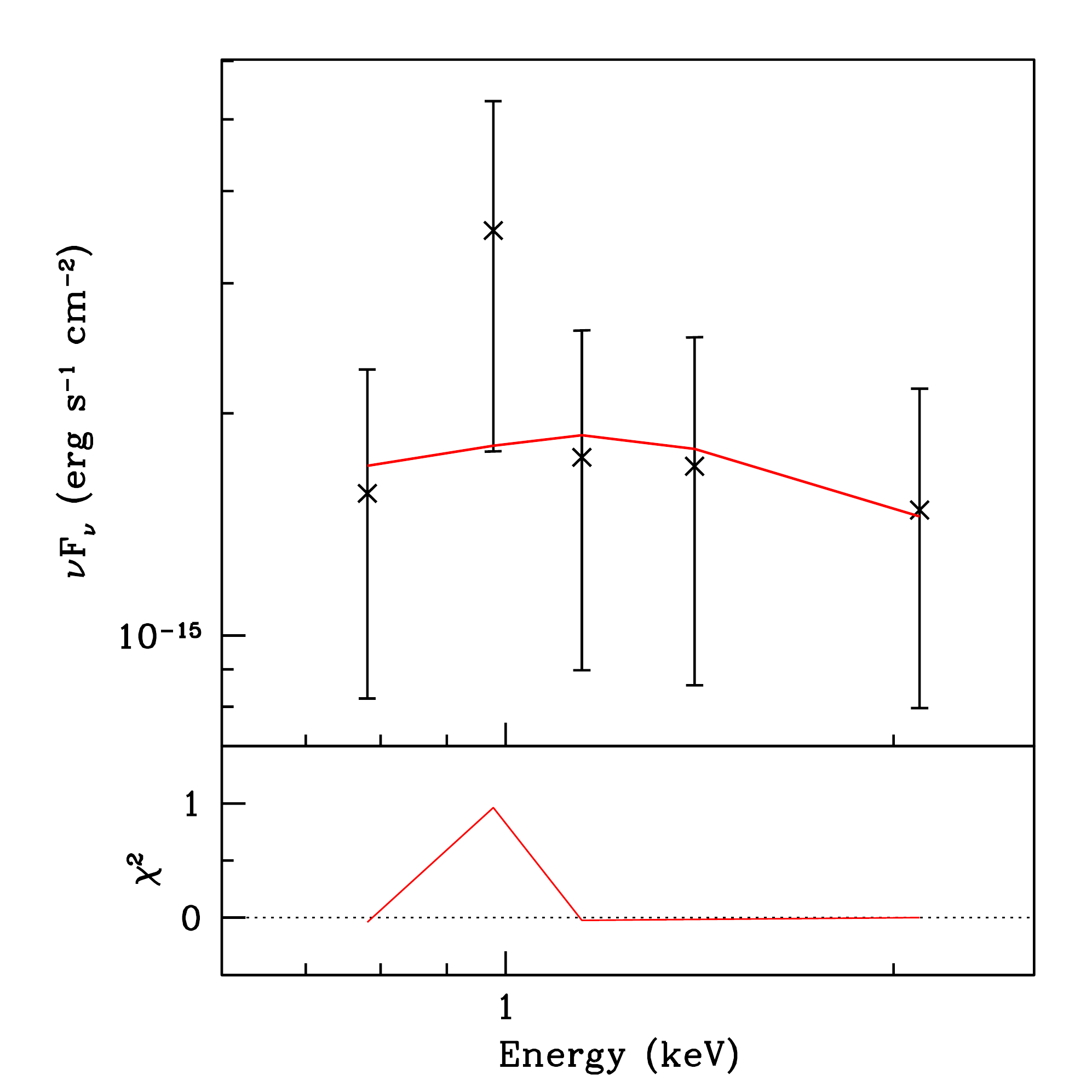} }
\qquad
\subfloat{ \includegraphics[width=0.45\textwidth]{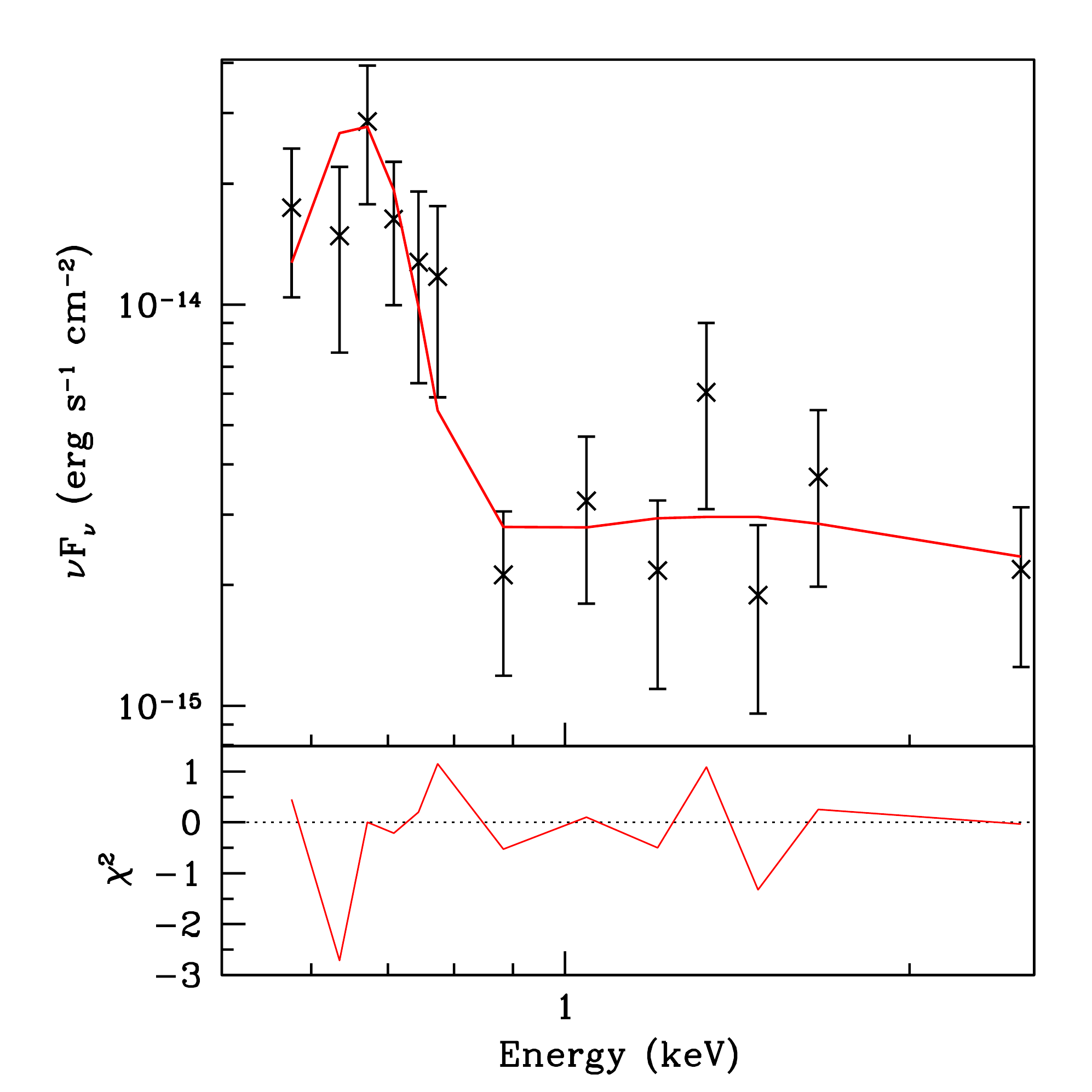} }
\caption[The bright and faint unfolded spectra of XMMU\,122939.7+075333 fit with continuum plus Gaussian model]{The bright and faint unfolded spectra of XMMU\,122939.7+075333 taken from Obs ID 12888 their respective best fit continuum plus Gaussian model (red line; see Table\,\ref{joint_fit} for spectral model fit parameters). The top left panel shows the faint Obs ID 12889 spectrum, the top right panel shows the faint Obs ID 12888 spectrum, the bottom left panel shows the bright Obs ID 12889 spectrum and the bottom right panel shows the bright Obs ID 12888 spectrum. The $\Delta$$\chi^2$ residual values are shown below each spectrum. }
\label{spectra}
\end{figure*}

\paragraph{Joint fits of Obs ID 12889, 12888, and XMM04 full spectra}
For the joint fit of Obs ID 12889, 12888 and XMM04  we compared the diskbb only, PL only, and diskbb plus PL continuum models with and without the Gaussian component. We then tested the significance of the line component using an F test. We note that these data fulfil the criteria for the appropriate use of the F test: the line energy is well-known in advance as the emission is assumed to be from O\,{\sc viii} and and the line width should be much smaller than the energy resolution of the CCDs on Chandra and XMM. So, while warnings about the use of the F test have been made in recent years \citep{2002ApJ...571..545P}, we are comfortable that it should provide, in this case, a good estimator of the significance of the feature.

\par
For the model consisting of a diskbb continuum only, the inner disk temperatures were 0.3$^{+0.1}_{-0.1}$\,keV, 0.5$^{+1.4}_{-0.2}$ and 0.3$^{+0.04}_{-0.04}$\,keV for Obs ID 12889, 12888, and XMM04 respectively. When the Gaussian component was added to the diskbb continuum model, the inner disk temperature, T$_{\rm in}$ for Obs ID 12889, 12888, and XMM04 were found to be 0.3$^{+0.2}_{-0.1}$\,keV, 0.7$^{+2.4}_{-0.4}$\,keV and 0.4$^{+0.1}_{-0.1}$\,keV respectively. An emission line component with a line energy of 0.66$^{+0.01}_{-0.04}$\,keV was found. In this fit, the line width was restricted to 0$-$0.04\,keV and a PW of 6.2$\times10^{-5}$\,keV was found. The EWs for Obs ID 12889, 12888, and XMM04 are 0.85\,keV, 0.17\,keV and 0.10\,keV respectively. The reduced $\chi^2$ is 0.88 for the diskbb only model and 0.76 for the Gaussian plus diskbb model. The F test revealed that the emission line is significant, with an F statistic of 8.24 and  null hypothesis probability of 4.6$\times 10^{-5}$.

\par
The PL-only continuum model yielded the following spectral parameters for Obs ID 12889, 12888, and XMM04 respectively: 3.5$^{+0.9}_{-0.6}$, 2.8$^{+1.5}_{-1.1}$ and 2.7$^{+0.2}_{-0.2}$. When the Gaussian component was included in the model, an emission line with an energy of 0.66$^{+0.02}_{-0.02}$\,keV and a PW of 1.92$\times10^{-6}$ $^{+0.04}_{-1.92\times10^{-6}}$\,keV was found. The spectral parameters for the continuum for this model are 3.10$^{+0.93}_{-0.83}$, $2.30^{+1.1}_{-0.8}$ and 2.5$^{+0.2}_{-0.2}$ for Obs ID 12889, 12888 and XMM04 respectively. The EWs are 0.32\,keV, 0.10\,keV and 0.07\,keV for Obs ID 12889, 12888, and XMM04 respectively. The PL-only model has a reduced $\chi^2$ of 0.73 whereas the Gaussian plus PL model had a reduced $\chi^2$ of 0.66. The F test results show that the line is significant; the F statistic is 5.91 and the null hypothesis probability is 8.22$\times10^{-4}$.

\par
In the Gaussian plus diskbb and PL continuum model, several of the continuum components were not well constrained, but the line parameters are robust and very similar to what is found for the other spectral fits. T$_{\rm in}$ for Obs ID 12889, 12888, and XMM04 were found to be 997.0$^{+>3.0}_{-996.0}$\,keV, 0.08$^{+0.03}_{-0.05}$\,keV, and 0.3$^{+0.003}_{-0.01}$\,keV respectively. The values for $\Gamma$ are 4.8$^{+>2.2}_{->6.8}$, 1.6$^{+0.2}_{-1.5}$, and 0.2$^{+2.1}_{-1.1}$. The Gaussian component had a line energy of 0.66$^{+0.02}_{-0.02}$\,keV and PW of 0.01$^{+0.03}_{->0.0}$\,keV. The EW values were 0.05$^{+0.03}_{-0.03}$\,keV, 0.11$^{+0.07}_{-0.07}$\,keV, and 0.06$^{+0.04}_{-0.04}$\,keV for Obs ID 12889, Obs ID 12888, and XMM04 respectively. The reduced $\chi^2$ for the Gaussian plus continuum model is 0.57. We obtain the reduced $\chi^2$ for the joint fit of the diskbb plus PL continuum model by summing the respective reduced $\chi^2$ values from the individual fits; this gives a reduced $\chi^2$ of 0.62. The F statistic is then 4.79 and the null hypothesis probability is 0.003. The line is therefore fairly significant.

\par
\begin{figure*}
 \centering
\subfloat{\includegraphics[width=0.4\textwidth]{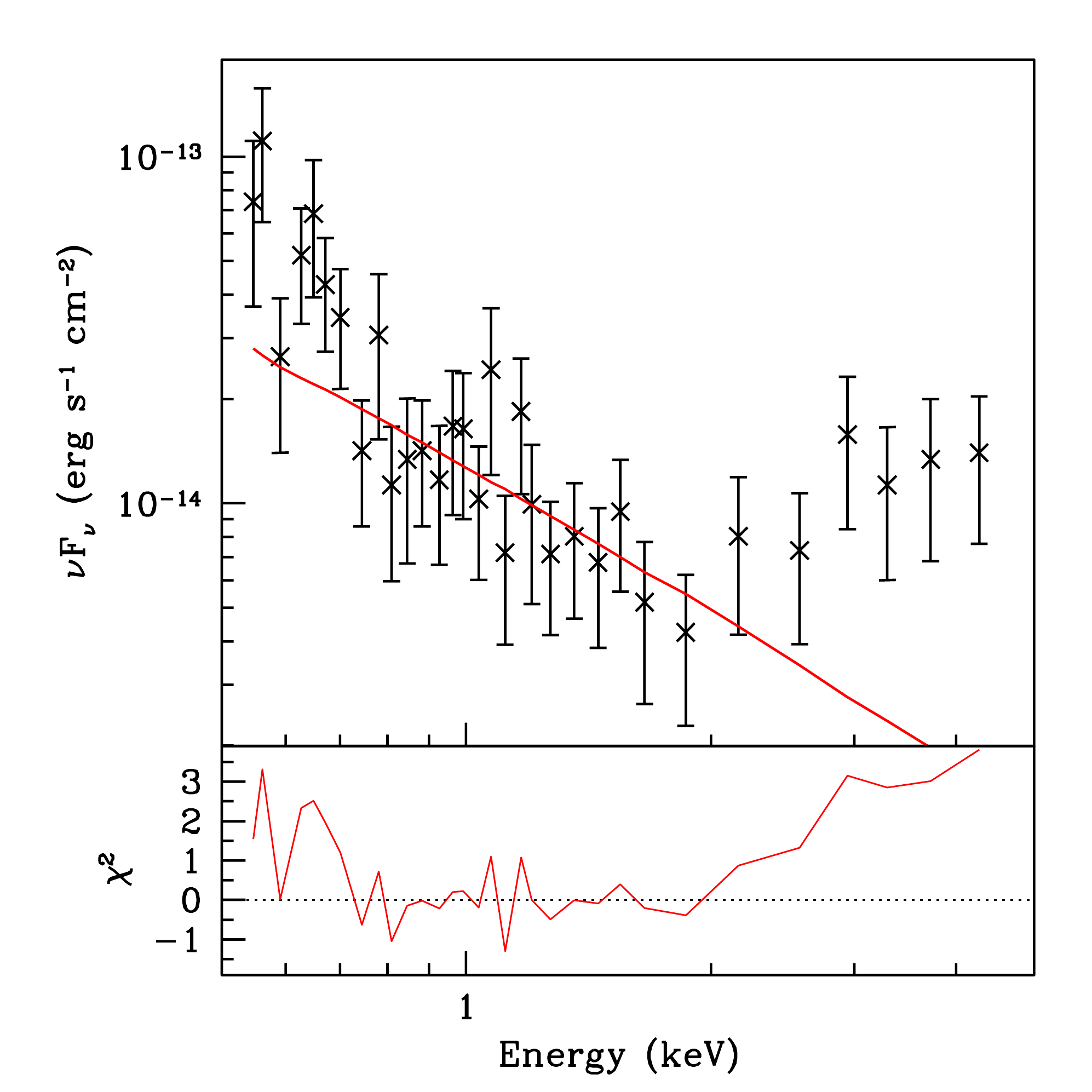}}
\subfloat{\includegraphics[width=0.4\textwidth]{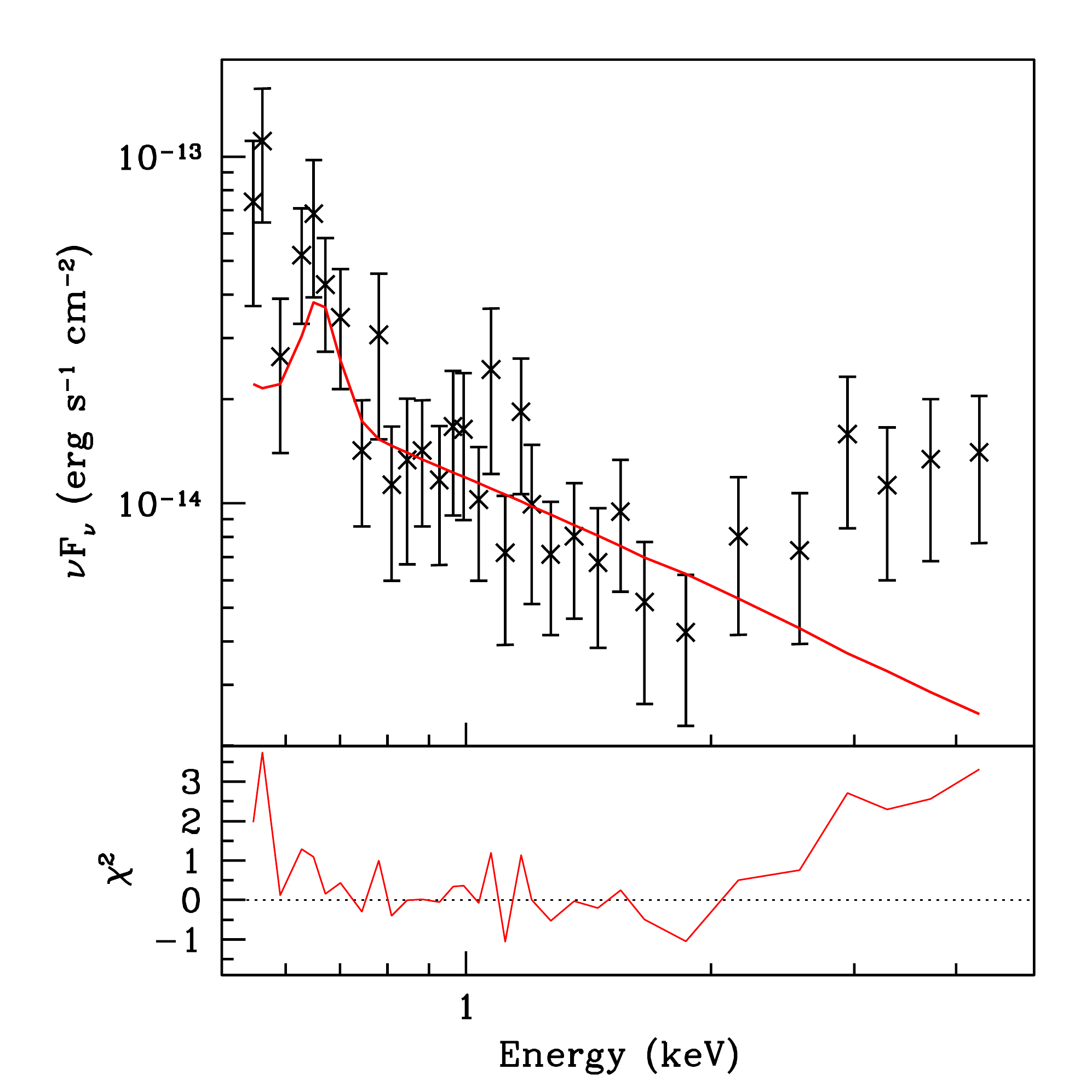} }
\qquad
\subfloat{\includegraphics[width=0.4\textwidth]{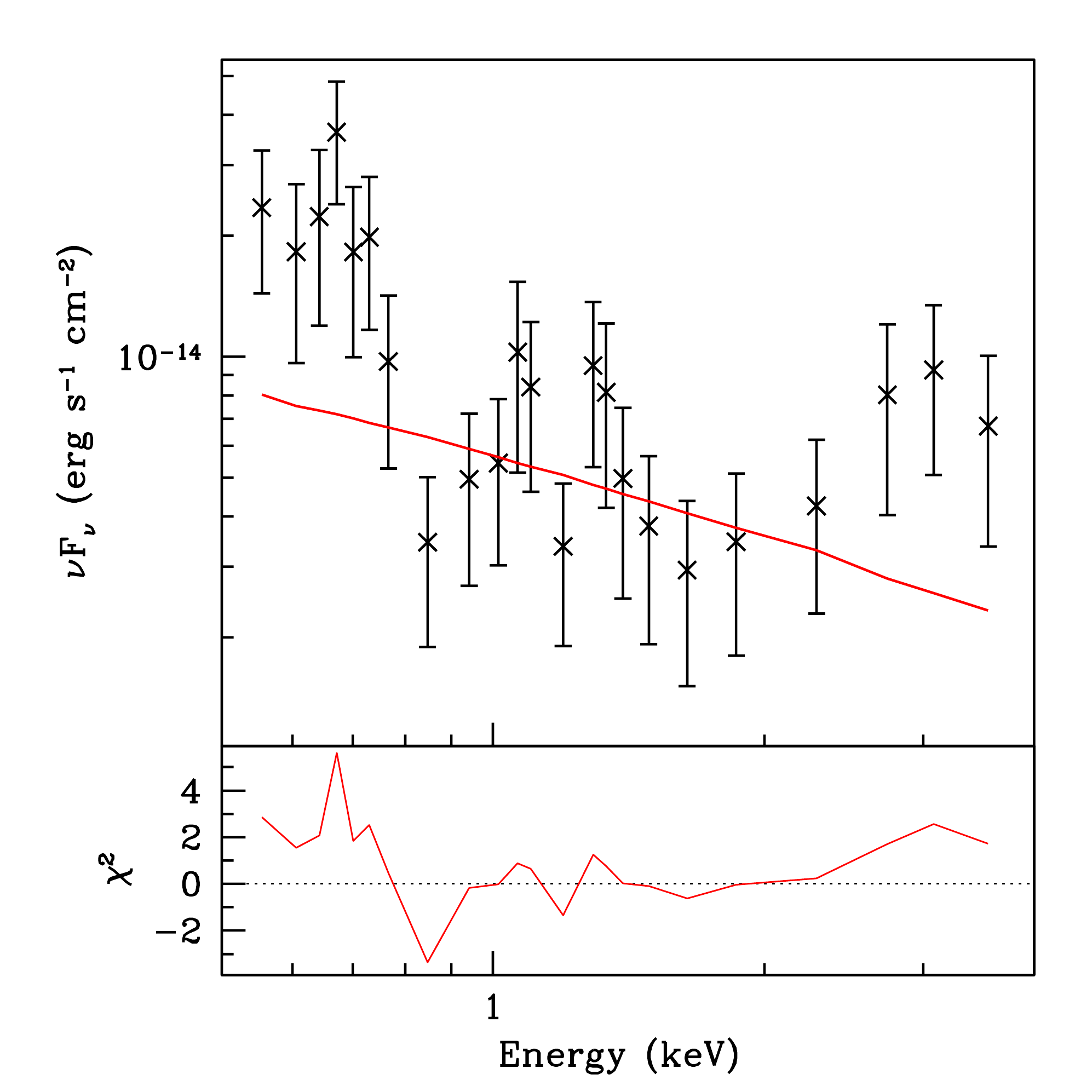}}
\subfloat{\includegraphics[width=0.4\textwidth]{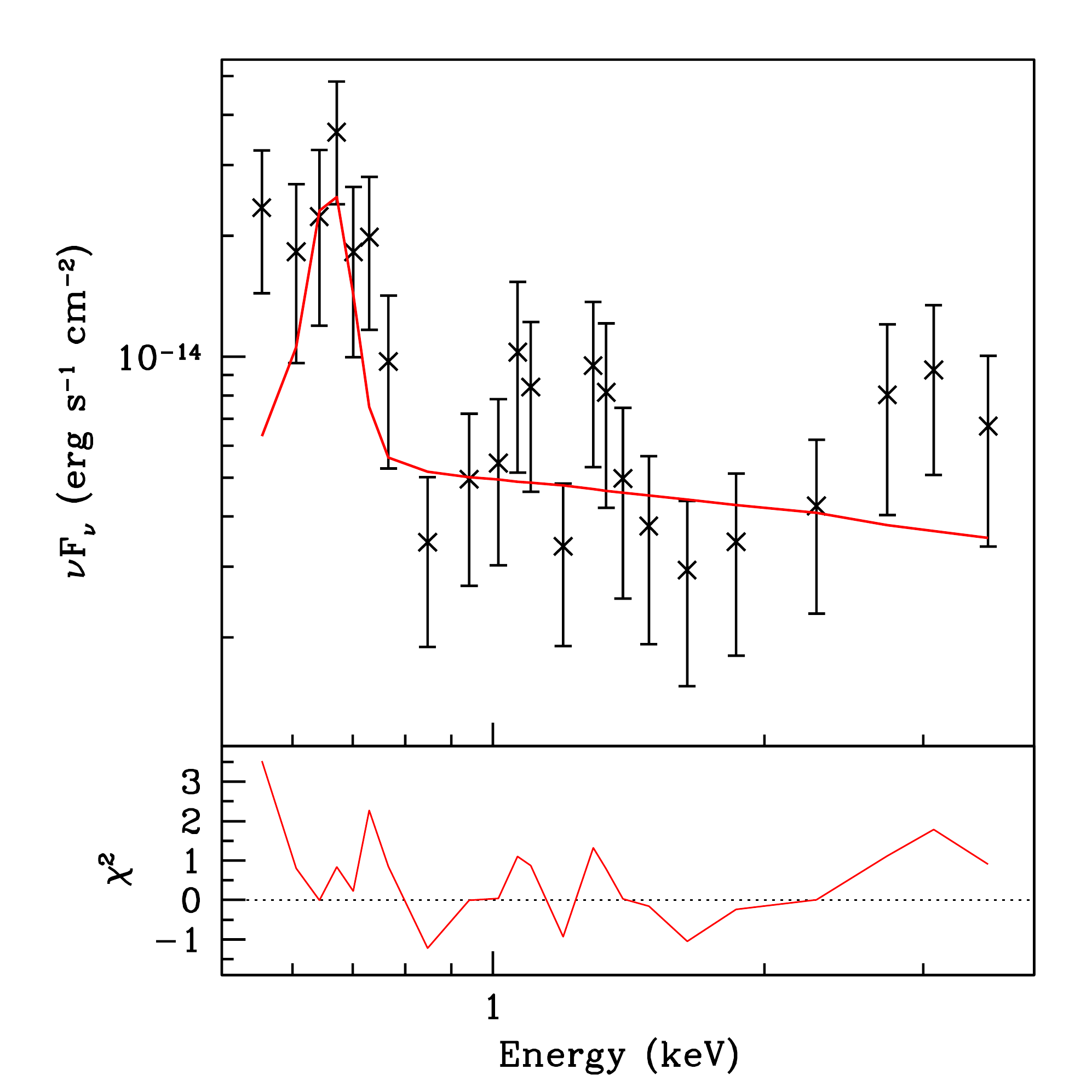} }
\qquad
\subfloat{\includegraphics[width=0.4\textwidth]{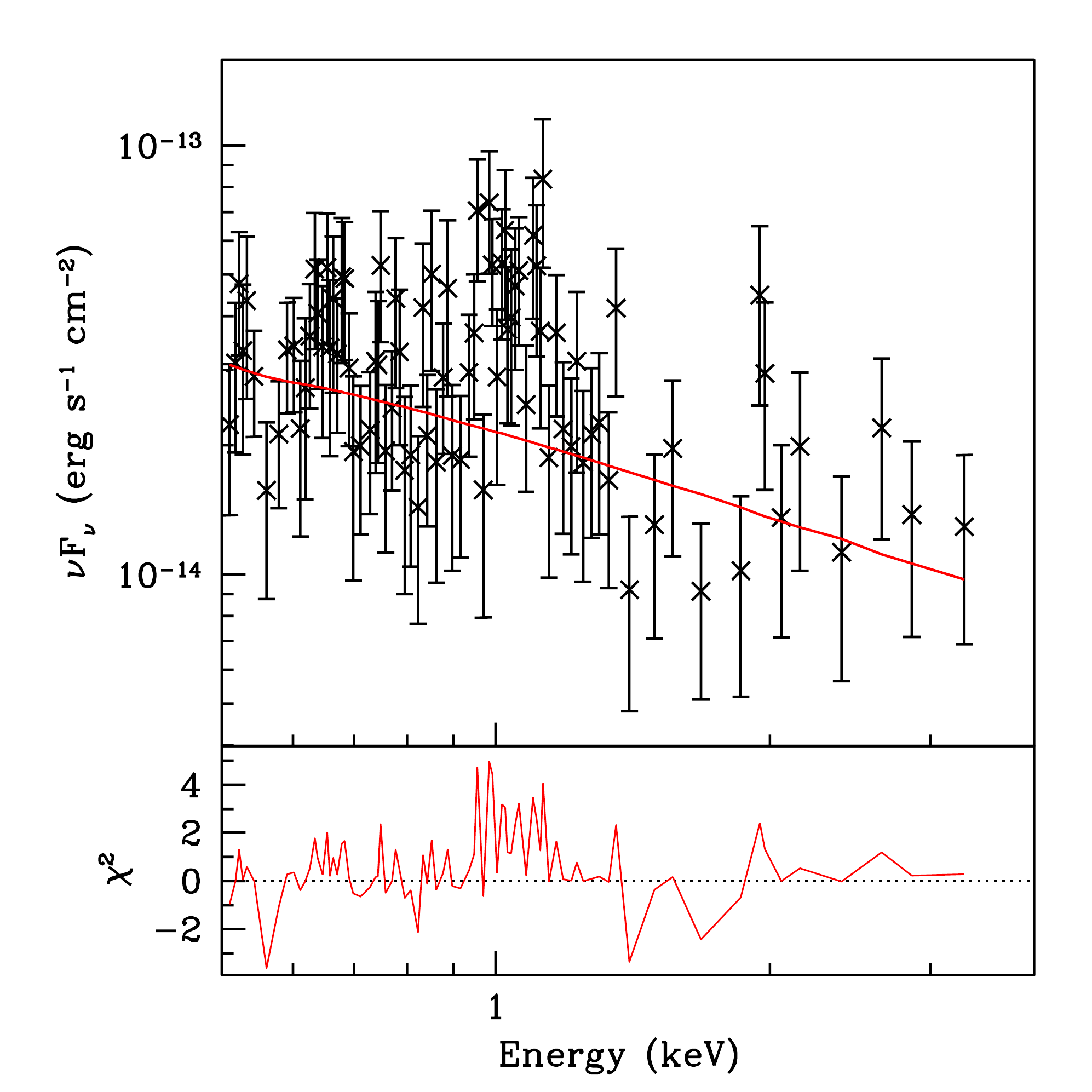}}
\subfloat{\includegraphics[width=0.4\textwidth]{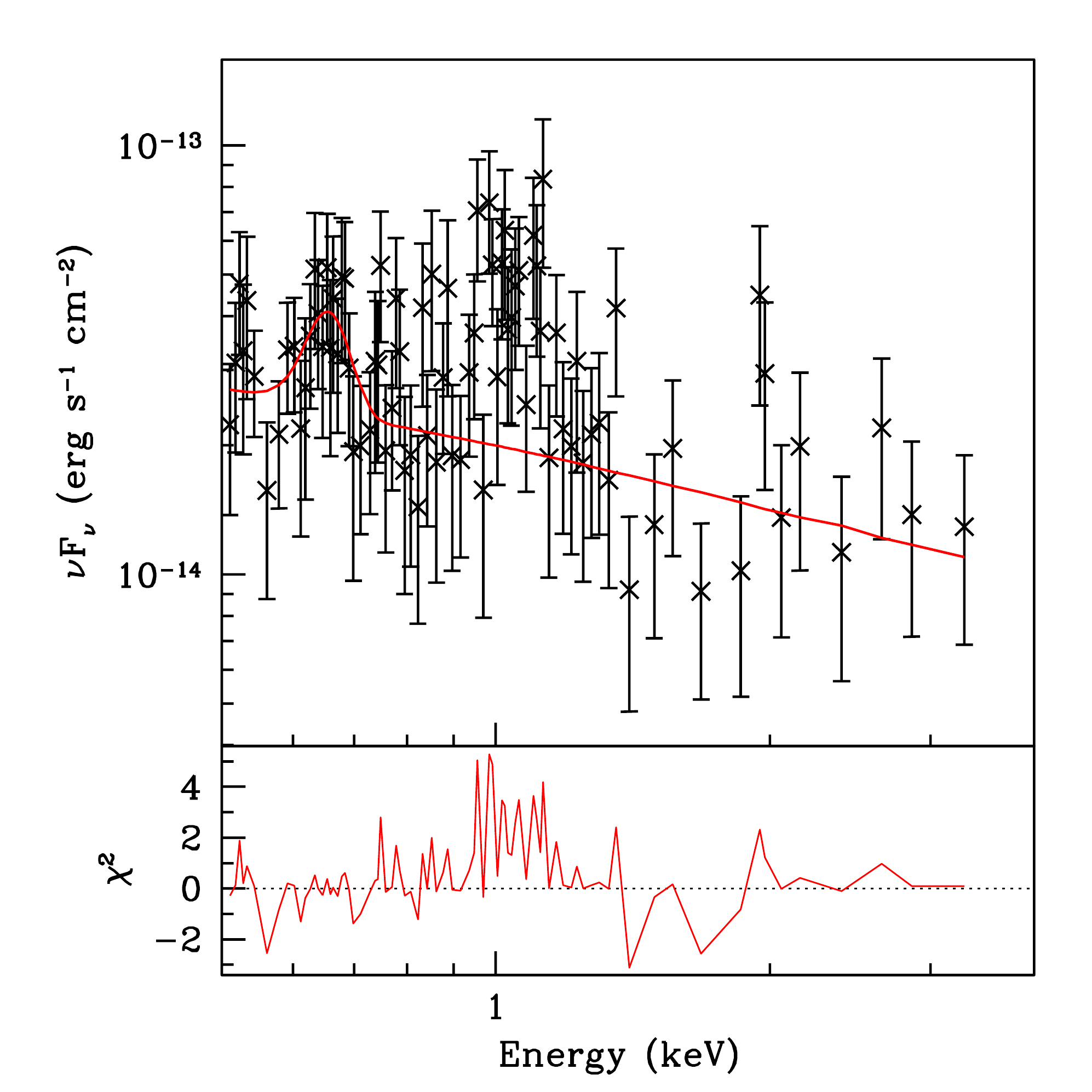} }
\caption[The unfolded spectra of XMMU\,122939.7+075333 fit with a PL continuum model]{The unfolded spectra of XMMU\,122939.7+075333 taken from Obs ID 12888 and 12889 and XMM04 with a PL continuum model (red line; see Table\,\ref{joint_fit} for spectral model fit parameters). The panels on the left hand side show spectra fit without the Gaussian component and the panels right show spectra fit with the Gaussian component. The top panels show the Obs ID 12889 spectra, the middle panels shows the Obs ID 12888 spectra and the bottom panels show the XMM04 spectra. The $\Delta$$\chi^2$ residual values are shown below each spectrum. }
\label{joint_fits_PL}
\end{figure*}
\begin{figure*}
 \centering
 \subfloat{\includegraphics[width=0.4\textwidth]{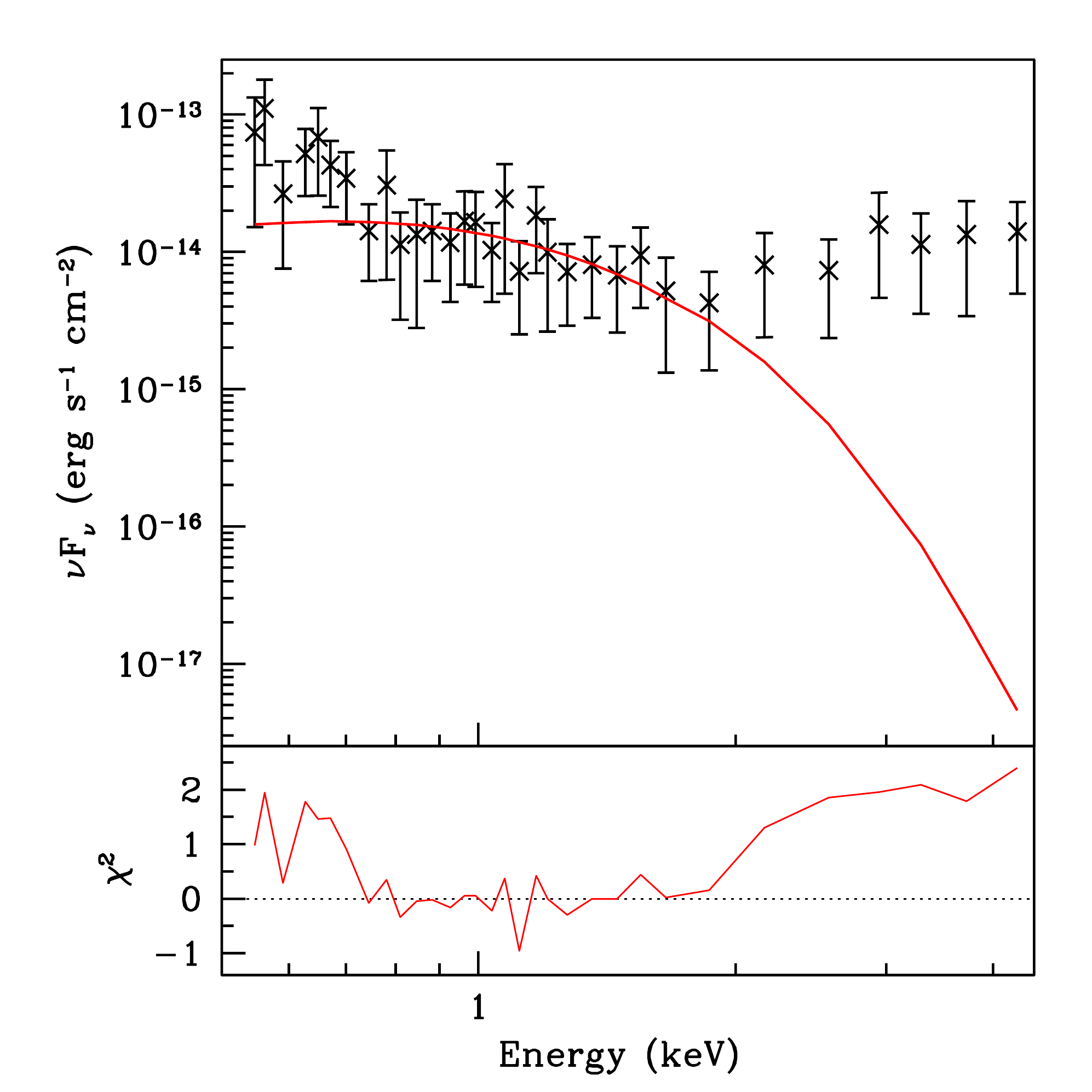}}
\subfloat{\includegraphics[width=0.4\textwidth]{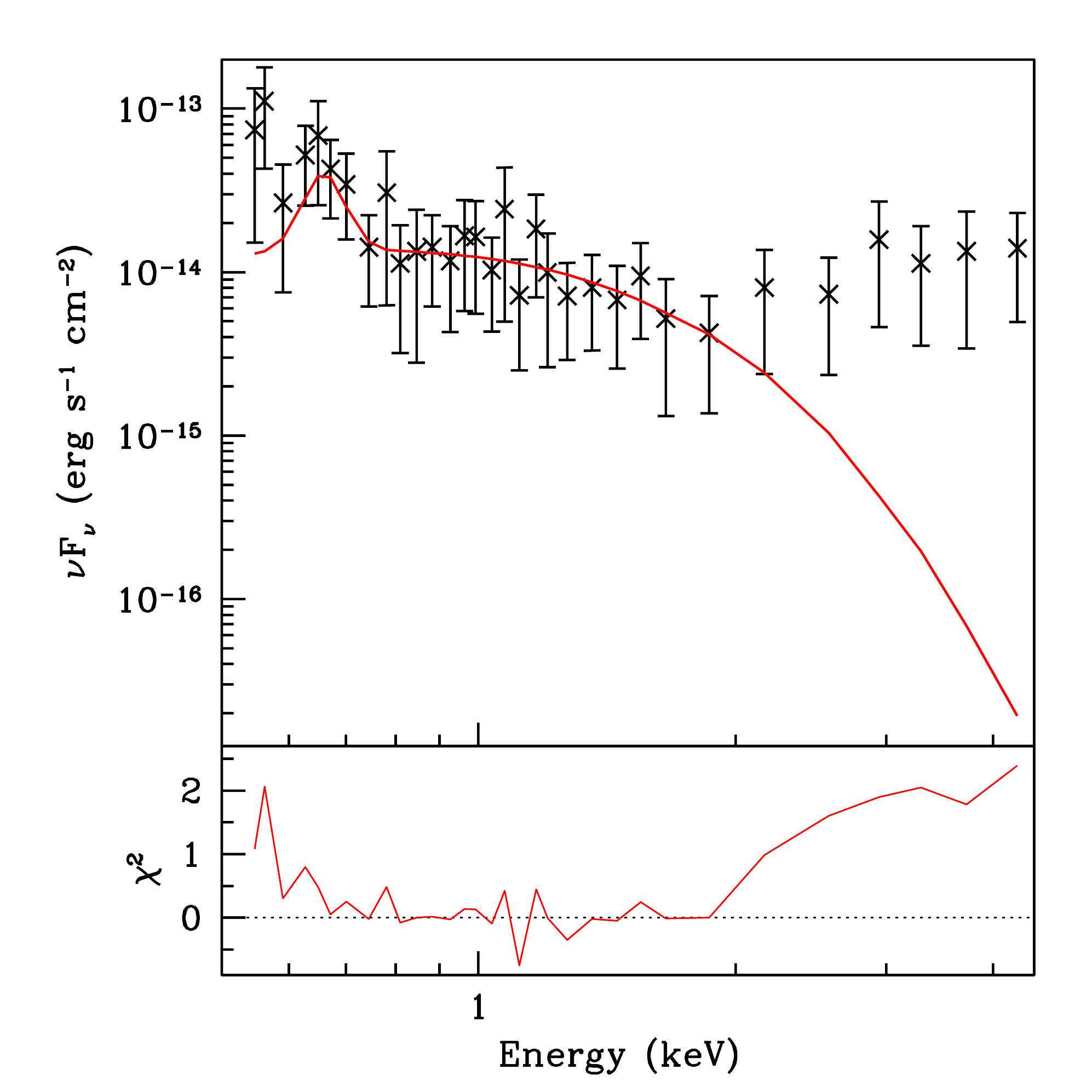} }
\qquad
\subfloat{\includegraphics[width=0.4\textwidth]{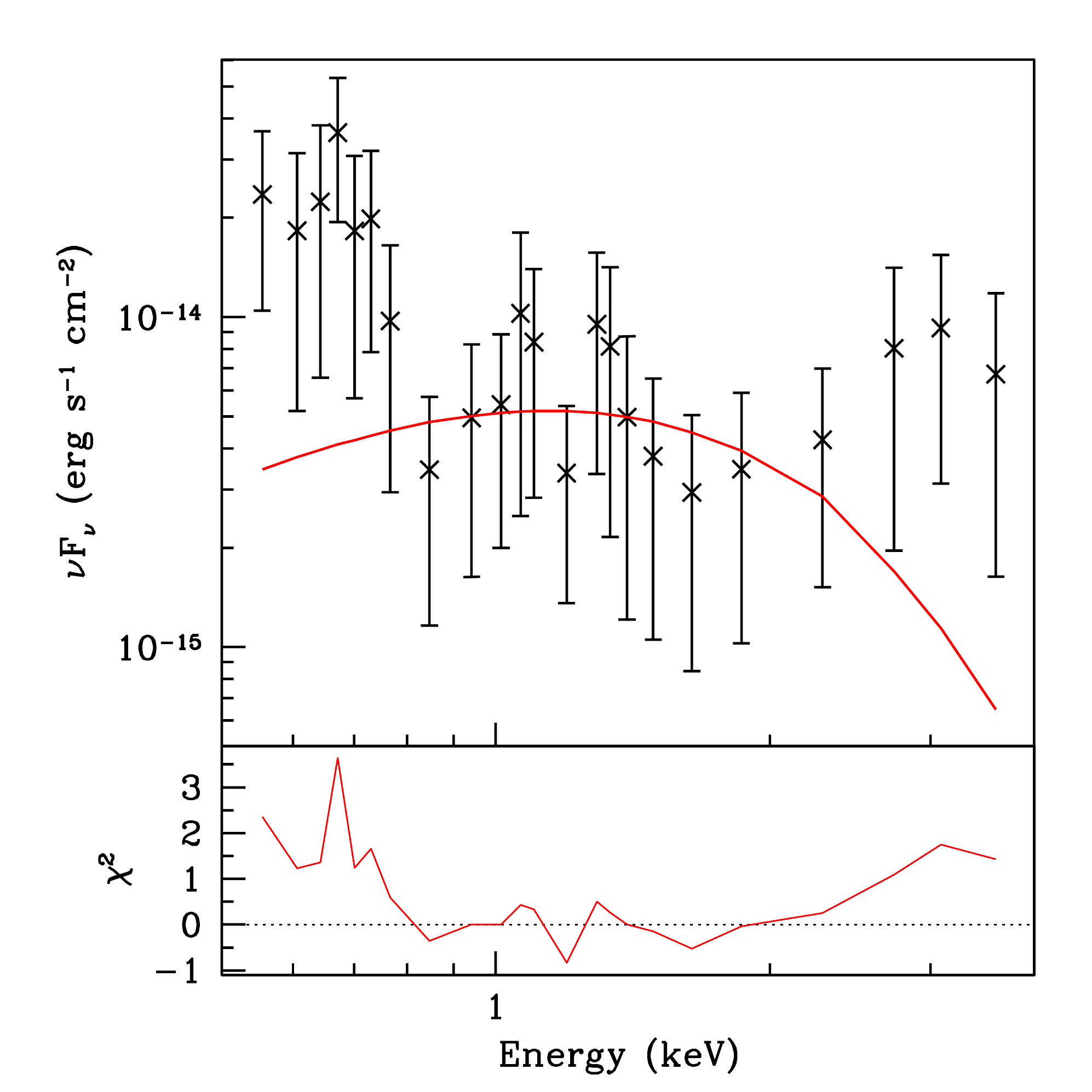}}
\subfloat{\includegraphics[width=0.4\textwidth]{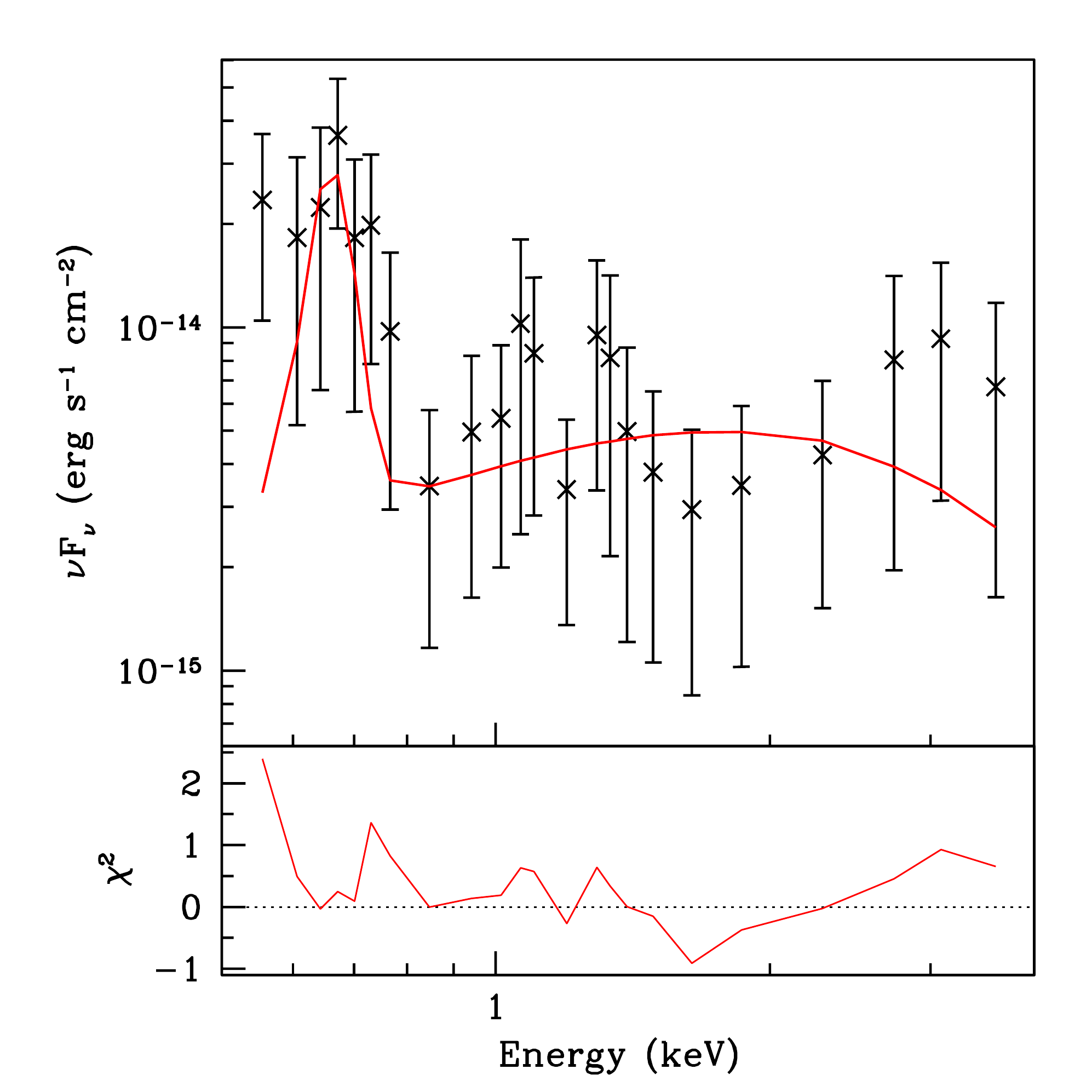} }
\qquad
\subfloat{\includegraphics[width=0.4\textwidth]{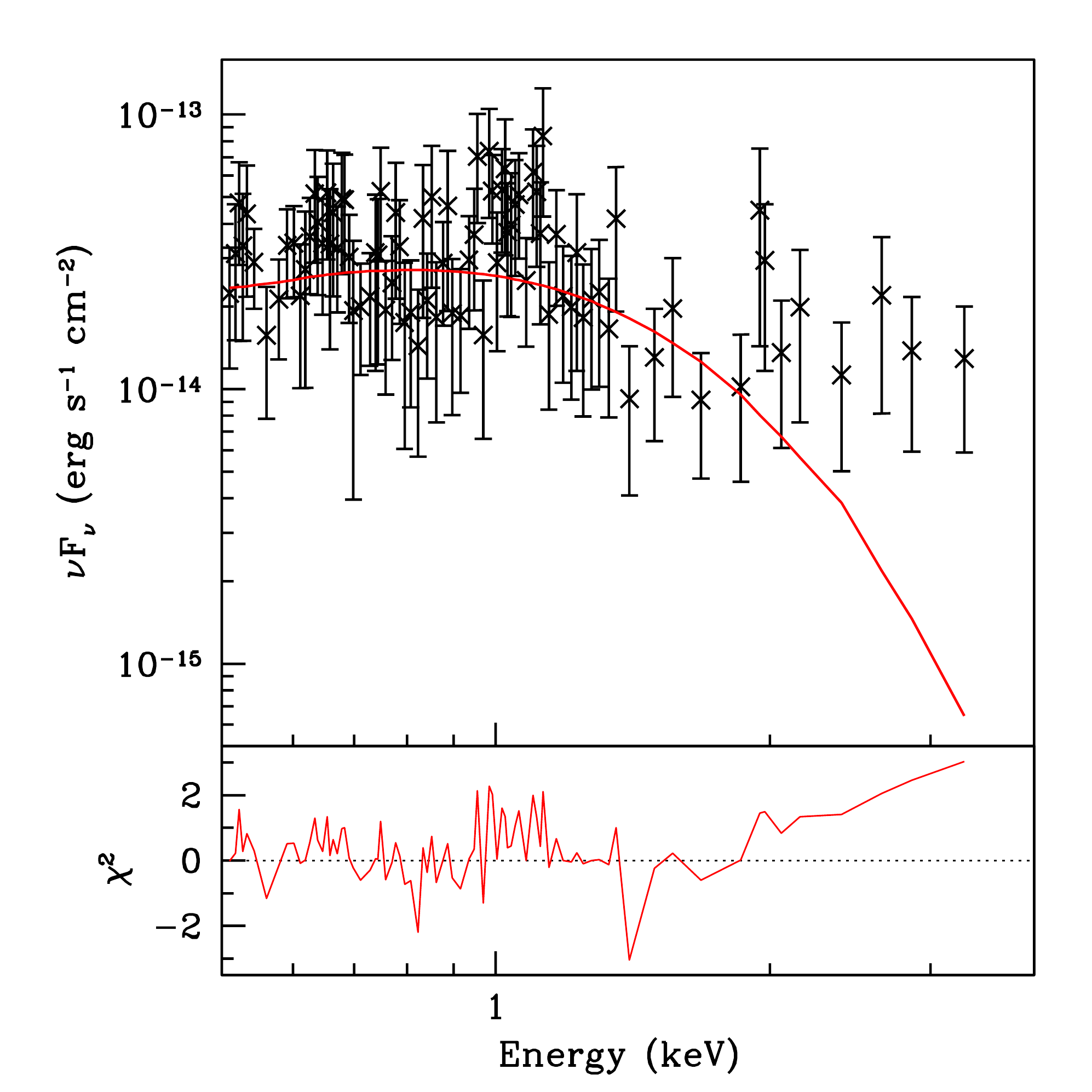}}
\subfloat{\includegraphics[width=0.4\textwidth]{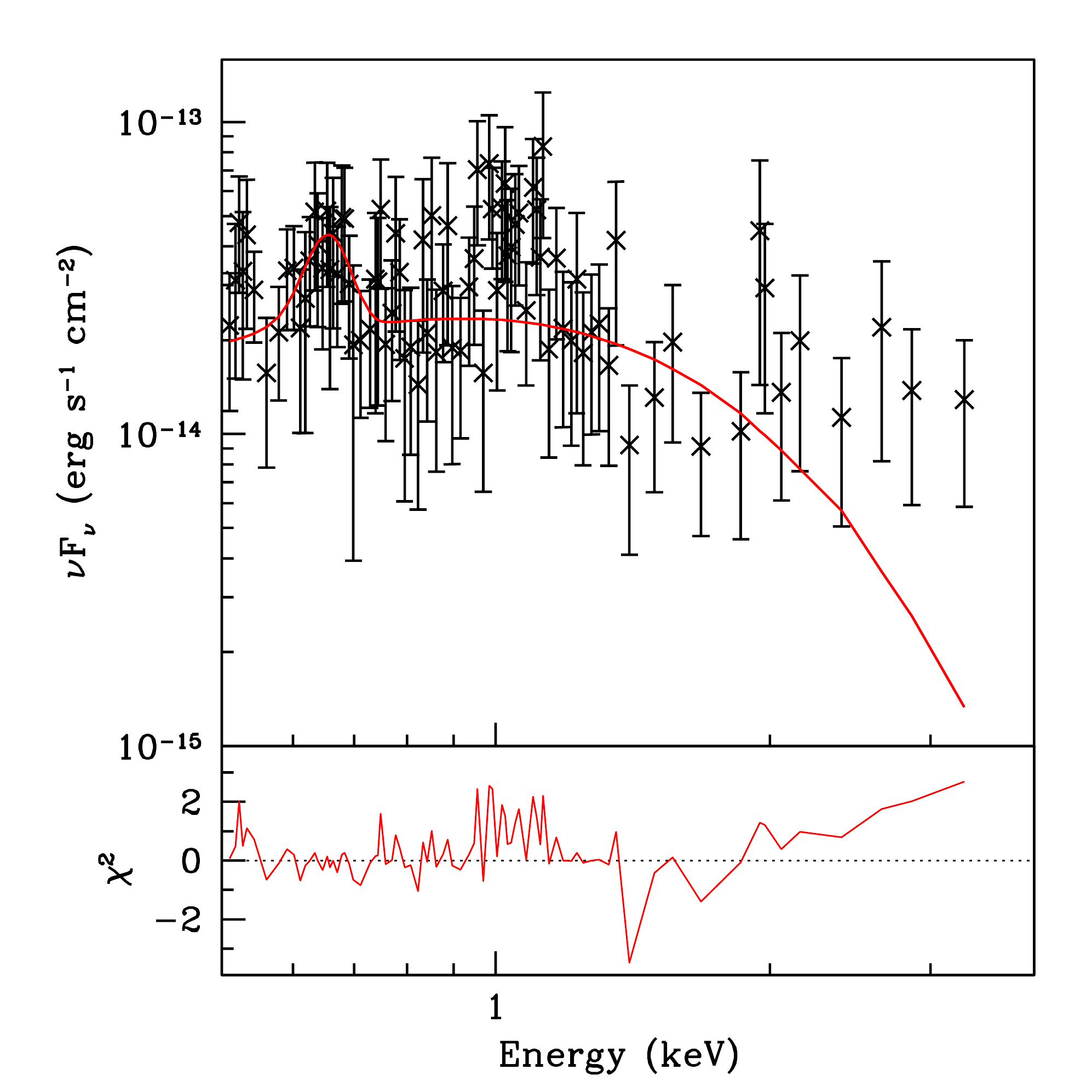} }
\caption[The unfolded spectra of XMMU\,122939.7+075333 fit with a diskbb continuum model]{The unfolded spectra of XMMU\,122939.7+075333 taken from Obs ID 12888 and 12889 and XMM04 with a diskbb continuum model (red line; see Table\,\ref{joint_fit} for spectral model fit parameters). The panels on the left hand side show spectra fit without the Gaussian component and the panels right show spectra fit with the Gaussian component. The top panels show the Obs ID 12889 spectra, the middle panels shows the Obs ID 12888 spectra and the bottom panels show the XMM04 spectra. The $\chi^2$ residual values are shown below each spectrum. }
\label{joint_fits_disk}
\end{figure*}

The spectral parameters for these models are are summarised in Table\,\ref{joint_fit}. The spectra for the diskbb only and PL only continuum model joint fits are shown in Fig.\,\ref{joint_fits_PL} and Fig.\,\ref{joint_fits_disk}.

\subsubsection{Varying absorption model}

We assume that the difference between the bright and faint phases in Obs ID 12888 are due to a variable absorption column density, i.e., the underlying continuum is essentially the same and the difference in flux is due to the soft photons being preferentially absorbed in the faint phase. We can test this assumption further by fitting the faint phase spectrum using the best fit continuum spectral parameters from the respective bright phase spectrum and leaving the absorption free to vary. If our assumption is correct, then the fits should be statistically acceptable and the absorption column density should be higher than the Galactic foreground value.

\par
We found that this model provided statistically acceptable fits for the faint spectrum, with $\chi^2$/$\nu=2.38/4$ for Obs ID 12888. The faint phase spectrum of Obs ID 12888 was found to have a higher absorption value of 2.5$\times10^{21}$\,cm$^{-2}$ and a 90\% confidence interval of $1.6\times10^{20} < N_{\rm H} < 9.6\times10^{21}$\,cm$^{-2}$.

\section{Discussion}
\subsection{Long term light curve}

We find that during the period from 1994 to 2011, the source has $L_{\rm X} \gtrsim 10^{38}$\,erg\,s$^{-1}$. In addition, the source luminosity has increased significantly between the observation of \citet{2010MNRAS.409L..84M} in February 2010 and the most recent observations in February 2011. This large increase in luminosity over a period of a year argues against the idea that the long term flux variations are caused by (just) the Kozai mechanism.

\subsection{Light curves of Obs ID 12889 and 12888}\label{short_lc_results}

This intra-observational  variability seen in Obs ID 12888 is too short to be due to spectral state changes, which have time scales on the order of days to weeks for stellar mass black hole binary systems \citep[see e.g.,][]{2009ApJ...701.1940Y}. These variations could, however, be easily explained by a variable absorption column density.

\subsection{Hardness Ratios}

This result is consistent with the explanation discussed in \S1: that a variable absorption column density is responsible for the decrease in X-ray luminosity due to the preferential absorption of soft X-ray photons. \citet{2008MNRAS.386.2075S} also found that the spectral changes between the faint and bright phases in XMM04 were due to a lack of soft photons in the faint phase, which is consistent with our result for Obs ID 12888.

\subsection{Spectral Analysis}

\subsubsection{Disk blackbody plus power law spectral model}\label{disk_PL}

\paragraph{Bright phase spectrum } 

The spectral parameters for the diskbb plus PL model are consistent with the system having a cool disk (T$_{\rm in}$=$0.08^{+0.06}_{-0.01}$\,keV) and hard power law component ($\Gamma$=2.06$^{+1.50}_{-1.42}$) as seen in other ULXs \citep[see e.g.,][]{2006MNRAS.368..397S}. This value of T$_{\rm in}$ is not consistent with the spectral parameter values found for Galactic black hole binaries, which are typically in the range 0.7 -- 1.5\,keV \citep[see e.g.][]{2006csxs.book..157M}.

\paragraph{Obs ID 12889 and 12888 and XMM04 full spectra}

These parameter values are in keeping with the results found for other ULXs; i.e. a cool disk (T$_{\rm in} \lesssim 0.2$\,keV) and a hard PL component ($\Gamma<1.5$) \citep[see e.g.,][]{2003ApJ...585L..37M,2005MNRAS.357..401J,2005MNRAS.357.1363R,2006MNRAS.368..397S}. The diskbb plus PL model has fit the apparent hard excess much better than the previous models that only include one continuum component (see Fig.\,\ref{disk_PL_spectra}). The disk blackbody plus power law model therefore provides a good fit to the spectra of XMMU\,122939.7+075333, implying that the source is behaving like a typical ULX. We note that the apparent excess of photons at around 1\,keV still remains, even with this two continuum component model. While this two component continuum model can fit ULX spectra with low signal-to-noise, \citep{2009MNRAS.397.1836G} showed that high quality ULX spectra with at least 10\,000 counts are not well fit by this model. Instead, these high signal-to-noise spectra are well fit a model consisting of a disc plus a Comptonized corona component. The characteristic disk temperature, T$_{\rm max}$,  for this model is found to be $\lesssim$0.5\,keV, which like the disk temperatures found for the diskbb plus PL models, is also significantly cooler than that that of the Galactic BHBs. Since very high quality spectra are required to properly fit the \citep{2009MNRAS.397.1836G} models, we do not attempt to carry out this analysis on our data sets.

\subsubsection{Continuum plus Gaussian spectral model}

\paragraph{Bright and faint phase spectra}

The energy for Gaussian component found in the models are consistent with emission from highly ionised oxygen, O\,{\sc viii}, with its line energy of 0.65\,keV. The O\,{\sc viii} edge absorption feature is located at 0.87\,keV. We have not attempted to model this feature due to the low signal-to-noise of the data, but cannot rule out its presence. However, because the fitted emission lines are strong and narrow, the absence or presence of the absorption edge should not affect whether the emission line is detected.

\paragraph{Obs ID 12889, 12888, and XMM04 full spectra}

Based on the results of the F test for these spectral fits, the Gaussian component is found to be fairly significant. In all the joint fits, the Gaussian component is narrow and unresolved. In the model with two continuum components, the Gaussian component is well constrained, with reasonable parameter values. However the continuum parameters in this model are degenerate with one another. Given that the disk blackbody component peaks outside of the \emph{Chandra} energy range, it is very difficult to distinguish between a cool diskbb and a steep PL continuum model.

\par
The emission line seen here is unlikely to be due to reflection from the accretion disk because it is so luminous \citep[see][who find an emission line with an equivalent width of 40\,eV from a Galactic ultracompact X-ray binary.]{2010MNRAS.407L..11M}. The emission could, however be caused by either collisional excitation or photoionisation. Due to the low signal-to-noise ratio of our data, we are not able to draw further conclusions about the origin of this spectral feature, except to say that this line emission seen in  XMMU\,122939.7+075333 and not in other nearby ULXs may be due to the much higher oxygen abundance in this system's donor star compared to systems fed by massive stars.

\par
These results suggest that the soft excess could be due to spectral line emission. We argue that this line emission could arise from highly ionised oxygen (O\,{\sc viii}), which is in line with our theory that the accretion disk of XMMU\,122939.7+075333 is rich in oxygen due to the companion star being a WD. 

\par
Other ULXs also show soft, line-like X-ray emission. For instance, \citet{2010MNRAS.402.2559C} found that the spectra of NGC\,5408 X-1 contained line-like emission at $\sim$0.6\,keV. They suggested that this emission could be due to the presence of highly ionised oxygen and iron around the source. Subsequently, \citet{2011MNRAS.411..644M} showed that this feature is well modelled by an optically thin plasma model component with temperature $\sim0.8$\,keV. NGC\,5204 X-1 was found to have an excess of counts at around 1\,keV \citep{2006MNRAS.371.1877R}. It was found that this excess could be modelled either as a thermal plasma component with $k {\rm T} \sim 0.9$\,keV or as a broad Gaussian line with $k {\rm T} \sim 0.96$\,keV. \citet{2006MNRAS.368..397S} found that the soft excess in NGC\,4395 X-1 could be modelled as a thermal plasma with $k{\rm T}=0.75$\,keV. 

\par
\citet{2014MNRAS.438L..51M} modelled the soft excess in NGC\,5408 X-1 and NGC\,6946 X-1 as blue-shifted, absorption lines due to the radiatively driven wind which is expected to accompany the accretion scenario in ULXs \citep{2011MNRAS.411..644M}. They find that modelling the spectra in this way gives statistically poorer fits than modelling with a thermal plasma. However, they still favoured broad absorption features from a disk wind over thermal plasmas, and argued that more sophisticated modelling is needed.  In our system, we argue that the features are more likely to be emission lines from the same plasma that produces the O\,[{\sc viii}] emission lines in the optical, but we agree that several plausible scenarios could produce the observed results, and some combination of better data and more sophisticated modelling would be required to distinguish between the various possibilities.

\par
Our spectra also show hints of other features besides the excess at 0.65\,keV. Both the Obs ID 12889 and 12888 spectra show suggestive evidence for a hard ($>$2\,keV) excess. \citet{2008MNRAS.386.2075S} also report a hard excess in their spectral analysis of the source. This feature may require an additional continuum component in order to fit properly. However, our data do not have enough counts to do this without increasing the degeneracy of the model significantly. There also appears to be an excess of photons at $\sim$1\,keV in the \emph{Chandra} and \emph{XMM} spectra. Narrow spectral features have been seen at these energies for LMC X-4 \citep{2009ApJ...696..182N,2010ApJ...720.1202H}. The narrow feature arises from emission from Ne X Ly${\alpha}$ (1.02 keV).

\par
However, since the standard ULX spectral model gives statistically acceptable fits to the spectra of XMMU\,122939.7+075333, we cannot say that the line plus continuum spectral modelled is preferred over the standard ULX model. We can therefore not distinguish between the scenario where the soft excess is due to highly ionised oxygen or a cool disk component, using these data sets.

\subsubsection{Varying absorption model}

The result of the varying absorption model indicate that the absorption column density is significantly higher during the faint phase of Obs ID 12888 than during the bright phase. This result is in accord with the findings of the flux variation analysis as well as the HR calculations. It therefore appears that over the course of Obs ID 12888, the source did undergo significant variations in flux, HR and absorption column density between the faint and bright phase. \citet{2008MNRAS.386.2075S}, using bright and faint phase spectra from XMM04, also found that the faint phase could be well modelled by using the best fit spectral parameters for the bright phase and letting the absorption column density be free to vary. They found a significantly increased value of $N_{\rm H}$ with this model.

\section{Conclusions}
In this paper we have shown that XMMU\,122939.7+075333 appears to be a persistently luminous X-ray source. The short term, intra-observational flux variability is too rapid to be caused by spectral state changes and can more easily be explained by a changing absorption column density. The significant increase in luminosity over a the period of a year (2010 to 2011) is not consistent with the idea that the long term, large amplitude flux modulations displayed by XMMU\,122939.7+075333 are caused by the Kozai mechanism alone. However, this does not rule out the possibility that there may be other variations in addition to the decay expected from the Kozai mechanism. For instance, \citet{2014MNRAS.439.1079A} have shown that periodic modulation of the eccentricity of a close binary can be forced by a third body, even in the absence of the large inclination angle required for Kozai resonances to develop.

\par
We also show that source has an excess of soft photons when the source is in a bright phase, which is consistent with previous analysis carried out on this system. The excess of soft X-ray photons can be modelled as emission from O\,{\sc viii} (0.65\,keV) from the source. However, the standard spectral model (disk blackbody plus power law component) also gives acceptable fits to the data. Both sets of spectral models suggest that XMMU\,122939.7+075333 is indeed BH binary. At this point we do not have sufficient evidence to prefer one over the other.

\section*{Acknowledgements}
TDJ acknowledges support from a Stobie-SALT studentship, funded jointly by the NRF of South Africa, the British Council and the University of Southampton. This research has made use of data obtained from the Chandra Data Archive and software provided by the Chandra X-ray Center (CXC) in the application packages CIAO and ChIPS, etc. This material is based upon work supported by the National Aeronautics and Space Administration under Grant No. GO1-12160X. GRS acknowledges the support of an NSERC Discovery Grant. TDJ would also like to thank Texas Tech University for its hospitality while this work was being finished. Lastly, the authors would like to thank Daniel Plant, Pablo Cassatella, Joey Neilsen, Matthew Middleton, and Dominic Walton for their assistance and useful comments. 


\bsp

\label{lastpage}

\end{document}